\documentclass[noeprint,twocolumn,aps,pra,showpacs,groupedaddress,superscriptaddress,nofootinbib]{revtex4-2}
\usepackage{graphicx}  
\usepackage[dvipsnames]{xcolor} 
\usepackage{color}     
\usepackage{amsmath}   
\usepackage{amsfonts}  
\usepackage{float}     
\usepackage[pdftex,linkcolor=blue,citecolor=blue,urlcolor=blue,colorlinks]{hyperref} 
\usepackage{mathalfa}
\usepackage[mathscr]{eucal}

\usepackage{tikz}
\usetikzlibrary{patterns} 
\usetikzlibrary{calc}
\usetikzlibrary{external}
\usetikzlibrary{arrows.meta}

\newcommand\varpm{\mathbin{\vcenter{\hbox{%
				\oalign{\hfil$\scriptstyle+$\hfil\cr
					\noalign{\kern-.3ex}
					$\scriptscriptstyle({-})$\cr}%
}}}}
\newcommand\varmp{\mathbin{\vcenter{\hbox{%
				\oalign{$\scriptstyle({+})$\cr
					\noalign{\kern-.3ex}
					\hfil$\scriptscriptstyle-$\hfil\cr}%
}}}}

\begin{document}

\title{Bosonic orbital Su-Schrieffer-Heeger model in a lattice of rings}

\author{Eul\`{a}lia Nicolau}
\affiliation{%
	Departament de F\'{i}sica, Universitat Aut\`{o}noma de Barcelona, E-08193 Bellaterra, Spain.
}%

\author{Anselmo M. Marques}%
\affiliation{Department of Physics and i3N, University of Aveiro, 3810-193 Aveiro, Portugal.}%

\author{Jordi Mompart}%
\affiliation{%
	Departament de F\'{i}sica, Universitat Aut\`{o}noma de Barcelona, E-08193 Bellaterra, Spain.
}%

\author{Ricardo G. Dias}%
\affiliation{Department of Physics and i3N, University of Aveiro, 3810-193 Aveiro, Portugal.}%

\author{Ver\`{o}nica Ahufinger}%
\affiliation{%
	Departament de F\'{i}sica, Universitat Aut\`{o}noma de Barcelona, E-08193 Bellaterra, Spain.
}%

\begin{abstract}
We study the topological properties of interacting and non-interacting bosons loaded in the orbital angular momentum states $l=1$ in a lattice of rings with alternating distances. At the single-particle level, the two circulation states within each site lead to two decoupled Su-Schrieffer-Heeger lattices with correlated topological phases. We characterize the topological configuration of these lattices in terms of the alternating distances, as well as their single-particle spectrum and topologically protected edge states. Secondly, we add on-site interactions for the two-boson case, which lead to the appearance of multiple bound states and edge bound states. We investigate the doublon bands in terms of a strong-link model and we analyze the resulting subspaces using perturbation theory in the limit of strong interactions. All analytical results are benchmarked against exact diagonalization simulations. 
\end{abstract}
\maketitle

\section{Introduction}\label{SecIntroduction}
A cornerstone idea behind topological insulators is the bulk-boundary correspondence. It relates the presence of robust edge states in a system with open boundary conditions with non-trivial values of topological invariants defined by the bulk bands. The symmetries and dimensionality of the non-interacting bulk restrict the possible topological phases that the system can host \cite{Qi2011,Chiu2016}. However, interacting systems do not possess a well-defined band structure with an associated topological invariant. In contrast to the characterization of non-interacting topological phases, a systematic description of interacting topological phases has yet to be developed \cite{Lin2022}.

The simplest case where interactions already play a role is the two-body problem. Repulsive and attractive interactions can cause the formation of bound pairs of particles with energies outside the non-interacting energy bands \cite{Hubbard1963,Mattis1986}. Such composite objects, usually called doublons \cite{Winkler2006,Creffield2004,Creffield2010}, have very long lifetimes due to the finite energy bandwidth of the single-particle kinetic energy \cite{Bello2017}. Doublons have been experimentally observed in ultracold atoms \cite{Winkler2006} and organic salts \cite{Wall2011}. Also, they have been shown to arise in a variety of systems, including models with long range interactions  \cite{Valiente2009,Valiente2010,Longhi2012,DiLiberto2017}, in superlattices \cite{Valiente2010a}, and in spinor gases \cite{Menotti2016}. 

Here we study a system of one or two bosons in a one-dimensional lattice of rings with alternating distances. This geometry mimics the Su–Schrieffer–Heeger (SSH) model \cite{Su1979,Zhang2003}, which was initially proposed to describe solitons in polyacetylene, and was latter revealed as the simplest instance of a topological insulator. Each local potential has eigenstates with orbital angular momentum (OAM) $l$ with winding numbers $\pm l$. The particles are loaded into the states with $l=1$ providing each site of the lattice with two internal states. The interacting two-particle SSH model was previously studied for both on-site and nearest-neighbor interactions \cite{DiLiberto2016,DiLiberto2017,Marques2017,Marques2018a}. Here, the additional degree of freedom in each site leads to a richer array of bound-states, edge bound states and strongly interacting subspaces. Ring potentials can be generated experimentally with a wide range of techniques (see \cite{Amico2021} and references therein), while the $l>0$ states can be excited using a rotating weak link \cite{Ramanathan2011,Wright2013}, by photon-to-atom OAM transfer 
\cite{Andersen2006,Franke-Arnold2017}, or by a temperature quench \cite{Corman2014a}. Alternatively, the physics described here can also be observed in the $p$ band of a conventional optical lattice \cite{Wirth2011,Li2016,Kiely2016,Kock2016}.

The rest of the article is organized as follows. In Sec.~\ref{SecPhysicalSystem}, we introduce the physical system and discuss the coupling strengths that appear between the different winding numbers. We analyze the single-particle case in Sec.~\ref{SecSingleParticle}, defining a basis rotation that decouples the system into two SSH chains that allow for a topological characterization of the system. We calculate their energy spectra and topologically-protected edge states for different distances.
In Sec.~\ref{SecTwoParticle}, we explore the two-boson case by introducing on-site Hubbard-like interactions in each site. We analyze the doublon bands in the energy spectrum in terms of a strong-link model. Additionally, we derive the effective Hamiltonians for the bound states in the regime of strong interactions, which lead to effective SSH and Creutz ladder models. Finally, we present our conclusions in Sec.~\ref{SecConclusion}.

\section{Physical system}\label{SecPhysicalSystem}  
\begin{figure}[t]
	\centering
	\includegraphics[width=1\columnwidth]{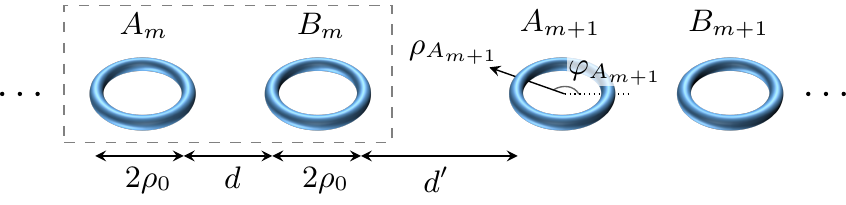}
	\caption{Representation of the considered one-dimensional lattice of rings. Each unit cell consists of two sites, $A_m$ and $B_m$, formed by identical ring potentials where $(\rho_{j_m},\varphi_{j_m})$ are the local radial and azimuthal coordinates at each site. The distance between adjacent sites alternates between $d$ for consecutive sites within a unit cell and $d'$ for the sites in adjacent unit cells.}
	\label{FigScheme}
\end{figure}

We consider bosons loaded into a one-dimensional lattice of ring potentials with alternating distances $d$ and $d'$. Each unit cell, $m$, includes the sites $A_m$ and $B_m$, as depicted in Fig.~\ref{FigScheme}, where we define the local polar coordinates for each site, $(\rho_{j_m},\varphi_{j_m})$ with $j=A,B$. The ring potential at each site is formed by a displaced harmonic potential in the radial coordinate, $V(\rho_{j_m})=\frac{1}{2} M \omega^{2}(\rho_{j_m}-\rho_0)^{2}$, where $\omega$ is the frequency of the radial potential, $M$, the mass of the atoms, and $\rho_0$, the radius of the ring. All the local potentials are identical, they have the same frequency $\omega$ and radius $\rho_0$. The distances $d^{(\prime)}$ are measured from one potential minima to the next, such that the distance that separates the unit cells is $D=d+d'+4\rho_0$ (see Fig.~\ref{FigScheme}).

\newpage 
The eigenstates of an isolated ring potential have well-defined OAM, $l$, and winding numbers $\nu=\pm l$. Each manifold of degenerate eigenstates with OAM $l$ is well separated in energy from the other manifolds, such that their dynamics become effectively decoupled in a lattice of rings \cite{Polo2016a,Pelegri2019}. The total field operator for a given OAM $l$ can be written as a linear combination of the local OAM eigenstates at each site of the lattice,
\begin{equation}\label{EqWavefunctionL}
	\begin{aligned}
		\hat{\Psi}_l(\mathbf{r})\hspace{-0.7mm}=\hspace{-0.5mm}& \sum_{m=1}^{N_{c}} \sum_{\nu=\pm l}\hspace{-0.8mm} \phi^{\nu}_{A_{m}}\hspace{-0.6mm}\left(\rho_{A_{m}}, \varphi_{A_{m}}\right) \hspace{-0.5mm}\hat{a}^{\nu}_{m}\hspace{-0.5mm}+\hspace{-0.5mm}\phi^{\nu}_{B_{m}}\hspace{-0.6mm}\left(\rho_{B_{m}}, \varphi_{B_{m}}\right) \hspace{-0.5mm}\hat{b}^{\nu}_{m}, 
	\end{aligned}
\end{equation}
where $N_c$ is the number of unit cells and $\hat{a}^{\nu}_{m}$ and $\hat{b}^{\nu}_{m}$ are the annihilation operators of the local OAM states $|j_m^\nu\rangle$, where $j=A,B$ denotes the site and $m$ labels the unit cell. We consider an integer number of unit cells $N_c$ throughout this work. The wavefunctions of each state $|j_m^\nu\rangle$ are given by
\begin{equation}
	\phi^{\nu}_{j_{m}}\left(\rho_{j_{m}}, \varphi_{j_{m}}\right)=\left\langle\mathbf{r} \mid j_{m}^ \nu\right\rangle=\psi\left(\rho_{j_{m}}\right) e^{i\nu\left(\varphi_{j_{m}}-\varphi_{0}\right)},
\end{equation}
where $\psi\left(\rho_{j_{m}}\right)$ is the radial part of the wavefunction and $e^{i\nu\left(\varphi_{j_{m}}-\varphi_{0}\right)}$ is the complex phase due to the non-zero OAM, where $\varphi_0$ indicates an arbitrary phase origin. 

The total Hamiltonian that describes the bosonic system is $\hat{\mathcal{H}}_l=\hat{\mathcal{H}}_{l}^{0}+\hat{\mathcal{H}}_{l}^{\mathrm{int}}$, with a single-particle Hamiltonian
\begin{equation}\label{EqKineticHamiltonian}
	\hat{\mathcal{H}}_{l}^0=\int d^2r\, \hat{\Psi}_{l}^{\dagger}(\mathbf{r})\left[-\frac{\hbar^{2} \nabla^{2}}{2 M}+V(\mathbf{r})\right] \hat{\Psi}_{l}(\mathbf{r}),
\end{equation}
where the potential $V(\mathbf{r})$ is the sum of the truncated harmonic potentials of each site, and an interaction term
\begin{equation}\label{EqInteractionHamiltonian}
	\hat{\mathcal{H}}^{int}_{l}=\frac{g}{2} \int d^2r\, \hat{\Psi}_{l}^{\dagger}(\mathbf{r}) \hat{\Psi}_{l}^{\dagger}(\mathbf{r}) \hat{\Psi}_{l}(\mathbf{r}) \hat{\Psi}_{l}(\mathbf{r}),
\end{equation}
where $g$ is proportional to the $s$-wave scattering length.

The tunnelling processes between OAM states of identical coplanar rings were thoroughly studied in \cite{Polo2016a} by analyzing the mirror symmetries of the single-particle case. The authors found that there are only three distinct tunnelling amplitudes that govern the dynamics: a $J_1$ term that couples the opposite circulation states within a single ring, a $J_{2}$ term that couples same circulation OAM modes in adjacent rings, and a $J_{3}$ term that couples opposite circulation modes in adjacent rings. Complex tunnelling amplitudes naturally arise in this system due to the nonzero OAM for certain geometries. However, for a lattice of inline rings, one can always choose the origin of the phase $\varphi_0$ along the lattice direction such that all couplings are real \cite{Polo2016a}.  

In this work, we study the states with OAM $l=1$ and winding numbers $\nu=\pm 1$, which we will denote by the circulation labels $\alpha=\pm$. The tunnelling strengths can be computed as follows. Consider a two-site lattice populated with the states $l=1$. The total field operator (\ref{EqWavefunctionL}) for this system includes the two circulation states for each ring, left and right. Thus, the dynamics of the system are well captured by a four-state model with four eigenvalues $\{E_i\}$. The tunnelling amplitudes can then be obtained from the eigenvalues as \cite{PelegriThesis}
\begin{equation}\label{EqCouplingsNumerical}
	\begin{aligned}
		J_1 & =\frac{1}{4}\left(E_1-E_2-E_3+E_4\right), \\
		J_2 & =\frac{1}{4}\left(-E_1+E_2-E_3+E_4\right), \\
		J_3 & =\frac{1}{4}\left(-E_1-E_2+E_3+E_4\right).
	\end{aligned}
\end{equation}
Each tunneling amplitude can be obtained for a particular ring separation $d$ by performing imaginary time evolution of the single-particle Hamiltonian (\ref{EqKineticHamiltonian}) on the two-ring system to find the exact eigenstates \cite{PelegriThesis}.  
The magnitudes of the couplings $|J_2|$ and $|J_3|$ decay with the separation $d$ between the two rings, and while $|J_3|>|J_2|$ for small values of $d$, they become equal for large distances \cite{Pelegri2019}. The coupling $|J_1|$, is approximately one order of magnitude smaller than the other two couplings for any distance $d$, so that it can be safely neglected in our analysis.

For the considered lattice of rings, the relevant tunnelling amplitudes are $J_2$ and $J_3$, which correspond to the intra-cell distance $d$, and $J_2'$ and $J_3'$, which correspond to the inter-cell distance $d'$ (see Fig.~\ref{FigScheme}). We introduce the total bosonic field operator (\ref{EqWavefunctionL}) in Eq.~(\ref{EqKineticHamiltonian}), use the above assumptions, and use harmonic oscillator units for the distances and energies, $\sigma=\sqrt{\hbar/M\omega}$ and $\hbar\omega$, respectively. Then, one arrives at the following single-particle Hamiltonian in terms of the creation and annihilation operators of the local OAM eigenstates
\begin{equation}\label{EqKineticHamiltonianHubbard}
\begin{aligned}\hat{\mathcal{H}}_{l=1}^0\hspace{-0.8mm}=&\hspace{0.8mm}J_2\sum_{m=1}^{N_c}\sum_{\alpha=\pm}\hat{a}_m^{\alpha\dagger}\hat{b}_{m}^{\alpha}+J'_2\sum_{m=1}^{N_c-1}\sum_{\alpha=\pm}\hat{b}_m^{\alpha\dagger}\hat{a}_{m+1}^{\alpha}+\\&\hspace{-0.6mm}+\hspace{-0.3mm}J_3\hspace{-0.6mm}\sum_{m=1}^{N_c}\sum_{\alpha=\pm}\hat{a}_m^{\alpha\dagger}\hat{b}_{m}^{-\alpha}\hspace{-0.3mm}+\hspace{-0.3mm}J'_3\sum_{m=1}^{N_c-1}\sum_{\alpha=\pm}\hat{b}_m^{\alpha\dagger}\hat{a}_{m+1}^{-\alpha}\hspace{-0.3mm}+\hspace{-0.3mm}\rm{H.c.}\end{aligned}
\end{equation}
We also consider on-site interactions, so that introducing the total bosonic field operator restricted to $l=1$ with $\nu=\pm1$ (\ref{EqWavefunctionL}) into Eq.~(\ref{EqInteractionHamiltonian}), one obtains \cite{PelegriThesis}
\begin{equation}\label{EqInteractionHamiltonianHubbard}
	\hat{\mathcal{H}}^{int}_{l=1}\hspace{-0.5mm}=\hspace{-0.5mm}\dfrac{U}{2}\hspace{-1mm}\sum_{j=a,b}\sum_{m=1}^{N_c}\!\!\left[ \hat{n}_{j_m}^+(\hat{n}^+_{j_m}\!\!-\!\!1)\!+\!\hat{n}^-_{j_m}(\hat{n}^-_{j_m}\!-\! 1)\!+\!4\hat{n}^+_{j_m}\hat{n}^-_{j_m}\right]\!,
\end{equation}
where $\hat{n}^{\alpha}_{j_m}=\hat{j}^{\alpha\dagger}_{m}\hat{j}^{\alpha}_{m}$ are the number operators for each site $j$, unit cell $m$ and circulation $\alpha$, and $U \equiv g \int d^2r\left|\psi\left(\rho_{j_{m}}\right)\right|^{4}$ is the interaction strength. Besides the conventional Bose-Hubbard like interaction terms for each circulation, there is an additional cross-circulation term with a greater strength.

\section{Single particle}\label{SecSingleParticle} 
\subsection{Band structure}
Let us explore the single particle case. One can obtain the Hamiltonian in momentum space by considering the limit $N_c\rightarrow\infty$ and expanding the creation and annihilation operators in the Hamiltonian (\ref{EqKineticHamiltonianHubbard}) as a Fourier integral 
\begin{equation}
\hat{j}^{\alpha}_{m}=\frac{1}{\sqrt{N_{c}}} \sqrt{\frac{D}{2 \pi}} \int_{-\frac{\pi}{D}}^{\frac{\pi}{D}} d k e^{-i k x_{m}} \hat{j}^{\alpha}_{k},
\end{equation}
where $\hat{j}^{\alpha}_{k}={\hat{a}^{\alpha}_{k},\hat{b}^{\alpha}_{k}}$, the distance that separates the unit cells is $D=d+d'+4\rho_0$, and $x_m$ is the position of the unit cell $m$. 
Then, the Hamiltonian in Eq. (\ref{EqKineticHamiltonianHubbard}) can be written in terms of the Hamiltonian in $k$ space as 
\begin{equation}
\hat{\mathcal{H}}_{l=1}^{0}=\oint_{B Z} \hat{\Psi}_{k}^{\dagger} \hat{\mathcal{H}}_{k} \hat{\Psi}_{k} d k,
\end{equation}
where $\hat{\Psi}_{k}^{\dagger}=\left(\hat{a}^{+\dagger}_{k}, \hat{a}^{-\dagger}_{k}, \hat{b}^{+\dagger}_{k}, \hat{b}^{-\dagger}_{k }\right)$. For our system, the Hamiltonian $\hat{\mathcal{H}}_{k}$ reads
\begin{equation}\label{EqHamiltonianKspace}
	\begin{pmatrix}
0 & 0 & J_2+J_2'e^{-ika} & J_3+J_3'e^{-ika}\\
0 & 0 & J_3+J_3'e^{-ika} & J_2+J_2'e^{-ika}\\
J_2+J_2'e^{ika} & J_3+J_3'e^{ika} & 0 & 0\\
J_3+J_3'e^{ika} & J_2+J_2'e^{ika} & 0 & 0
\end{pmatrix}\!\!,\end{equation}
and has the following eigenvalues:
\begin{equation}\label{EqEnergyBands}
	\begin{aligned}
		\epsilon_{1,2}&=\pm\sqrt{t_a^{\prime2}+t_a^2+2t_a^\prime t_a\cos{(ka)}},\\
		\epsilon_{3,4}&=\pm\sqrt{t_s^{\prime2}+t_s^2+2t_s^\prime t_s\cos{(ka)}},
	\end{aligned}
\end{equation}
where 
\begin{equation}\label{EqCouplingsSSH}
	\begin{aligned}
		t_a^\prime=J_2'-J_3', \qquad & t_a=J_2-J_3, \\
		t_s^\prime=J_2'+J_3', \qquad & t_s=J_2+J_3.
	\end{aligned}
\end{equation}
The system presents two sets of energy bands $\epsilon_{1,2}$ and $\epsilon_{3,4}$ that are symmetrical with respect to zero energy. The energy bands $\epsilon_1$ and $\epsilon_2$ tend to degeneracy at zero energy for large intertrap separations $d^{(\prime)}$, for which $J_2^{(\prime)}=J_3^{(\prime)}$.  A direct topological characterization of the system does not respect the bulk-boundary correspondence, due to the presence of a unitary symmetry defined by the exchange of circulations $+\leftrightarrow -$ in each site, which we will discuss later. In the next section we perform a basis rotation that leaves the Hamiltonian in block diagonal form, allowing for a topological characterization of the system within each block.

\begin{figure}[t]
	\centering
	\includegraphics[width=1\columnwidth]{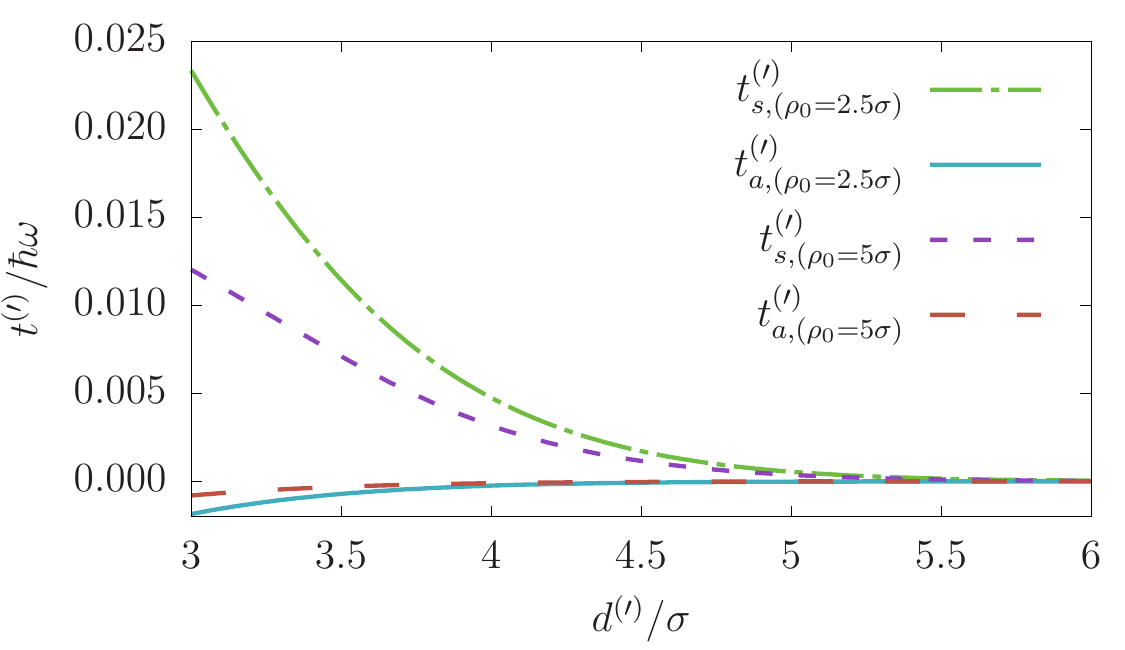}
	\caption{Couplings $t_a^{(\prime)}$ and $t_s^{(\prime)}$ as a function of the separation distance $d^{(\prime)}$ between rings for $\rho_0=2.5\sigma$ and $\rho_0=5\sigma$, where $\sigma=\sqrt{\hbar/M\omega}$ is the harmonic oscillator length.}
	\label{FigCouplingsVsDistance}
\end{figure}

\subsection{Mapping into two decoupled SSH chains}
We consider the symmetric ($s$) and antisymmetric ($a$) superpositions of the positive and negative circulations in each site
\begin{equation}\label{EqBasisChange}
	\begin{aligned}&\Big|A_{m}^{s(a)}\Big\rangle=\frac{1}{\sqrt{2}}\left(\left|A_{m}^+\right\rangle \varpm\left|A_{m}^-\right\rangle\right),\\
		&\Big|B_{m}^{s(a)}\Big\rangle=\frac{1}{\sqrt{2}}\left(\left|B_{m}^+\right\rangle \varpm\left|B_{m}^-\right\rangle\right).\end{aligned}
\end{equation}
In this new basis, the single particle Hamiltonian in (\ref{EqKineticHamiltonianHubbard}) can be block diagonalized into two decoupled SSH chains. The symmetric chain is described by the Hamiltonian
\begin{equation}\label{EqSSHHamiltoniansSym}
\begin{aligned}
\hat{\mathcal{H}}_s=\,&t_s\sum_{m=1}^{N_c}\hat{a}^{s\dagger}_{m}\hat{b}^s_{m}+t_s'\sum_{m=1}^{N_c-1}\hat{a}^{s\dagger}_{m+1}\hat{b}^s_{m}+\mathrm{H.c.},\\
\end{aligned}
\end{equation}
where $\hat{a}^s_m$ and $\hat{b}^s_m$ are the annihilation operators of the symmetric states defined in Eq.~(\ref{EqBasisChange}) and the couplings $t_s^{(\prime)}$ (\ref{EqCouplingsSSH}) define the energy bands $\epsilon_{3,4}$ of Eq.~(\ref{EqEnergyBands}). Similarly, the antisymmetric chain is described by the Hamiltonian
\begin{equation}\label{EqSSHHamiltoniansAsym}
\begin{aligned}
\hat{\mathcal{H}}_a=\,&t_a\sum_{m=1}^{N_c}\hat{a}^{a\dagger}_{m}\hat{b}^a_{m}+t_a'\sum_{m=1}^{N_c-1}\hat{a}^{a\dagger}_{m+1}\hat{b}^a_{m}+\mathrm{H.c.},\\
\end{aligned}
\end{equation}
where $\hat{a}^a_m$ and $\hat{b}^a_m$ are the annihilation operators of the antisymmetric states defined in Eq.~(\ref{EqBasisChange}) and the couplings $t_a^{(\prime)}$ (\ref{EqCouplingsSSH}) define the energy bands $\epsilon_{1,2}$ of Eq.~(\ref{EqEnergyBands}). Thus, each SSH chain contributes two energy bands to the whole system. 

Figure \ref{FigCouplingsVsDistance} shows the couplings $t_a^{(\prime)}$ and $t_s^{(\prime)}$ as a function of the separation distance between two rings $d^{(\prime)}$ obtained numerically using Eqs.~(\ref{EqCouplingsNumerical}) and (\ref{EqCouplingsSSH}). We represent the cases of two ring radii, $\rho_0=2.5\sigma$ and $5\sigma$, where $\sigma=\sqrt{\hbar/M\omega}$ is the harmonic oscillator length. All couplings decay with the separation distance, but due to the dependence of $t_s^{(\prime)}$ and $t_a^{(\prime)}$ on the couplings $J_2^{(\prime)}$ and $J_3^{(\prime)}$, $t_a^{(\prime)}$ remains much smaller than $t_s^{(\prime)}$ regardless of the ring radius $\rho_0$ and frequency $\omega$. 

\begin{figure}[t]
	\centering
	\includegraphics[width=0.438\columnwidth,trim=0cm 0cm 1.45cm 0cm,clip]{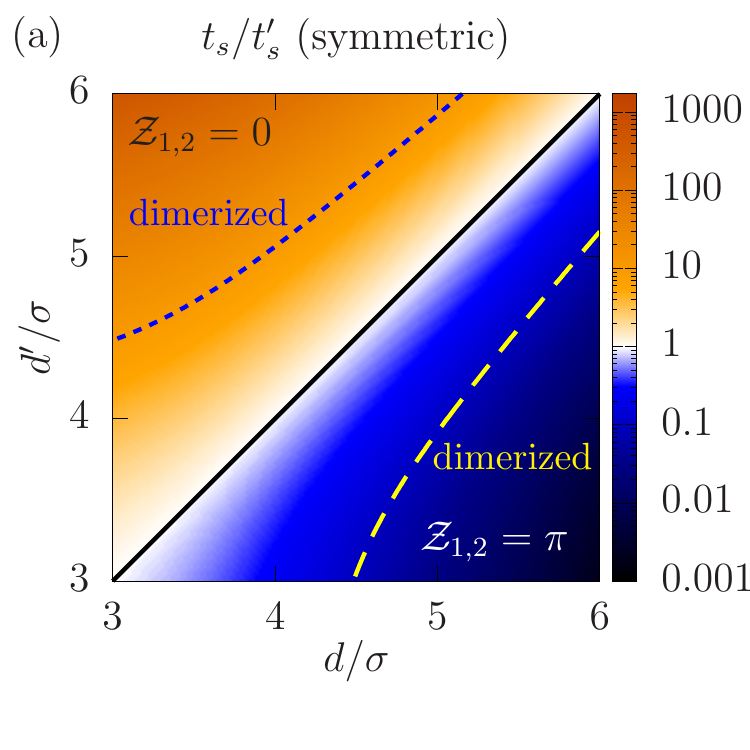}
	\includegraphics[width=0.542\columnwidth,trim=0cm 0cm 0cm 0cm,clip]{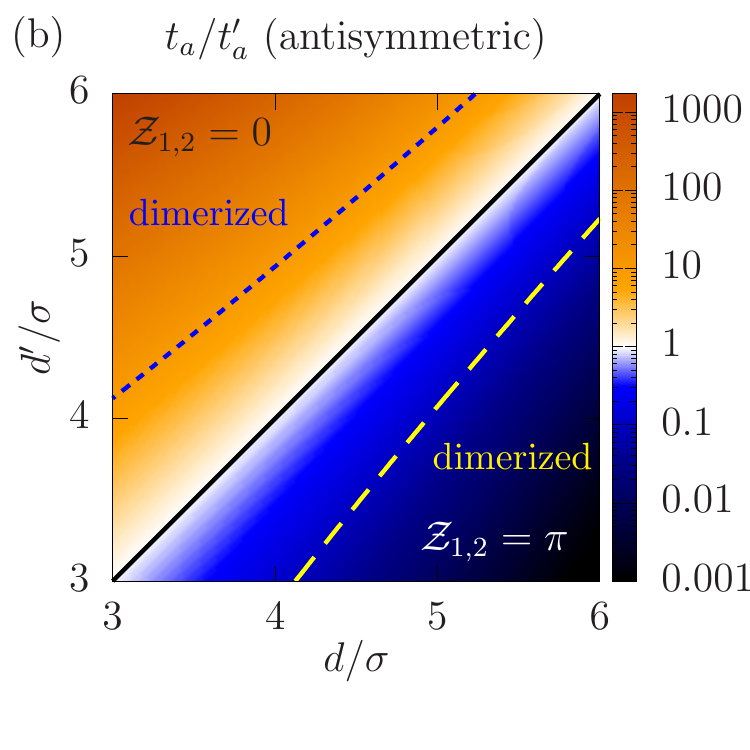}
	\caption{Phase diagram of the (a) symmetric, $\hat{\mathcal{H}}_s$, and (b) antisymmetric, $\hat{\mathcal{H}}_a$, chains as a function of the separation distances between rings, $d^{(\prime)}$, for  $\rho_0=5\sigma$, where $\sigma=\sqrt{\hbar/M\omega}$ is the harmonic oscillator length. Color represents the ratios (a) $t_s/t_s'$ and (b) $t_a/t_a'$, the black line separates the trivial phases with $\mathcal{Z}_{1,2}=0$ (warm colors) and the topological phases with $\mathcal{Z}_{1,2}=\pi$ (cold colors). The dashed blue lines and dotted yellow lines bound the nearly dimerized regimes of the trivial and topological phases, respectively.}
	\label{FigTopologicalPhases}
\end{figure}

In the SSH model \cite{Asboth2015}, the topological phase of the system is determined by the ratio of the couplings, $t/t'$, which determines the value of the Zak phase in each energy band $p$, 
\begin{equation}
\mathcal{Z}_p=i \oint_{B Z}\big\langle u_{k}^{p}\big|\partial_{k}\big| u_{k}^{p}\big\rangle \, d k,
\end{equation}
where $|u_k^p\rangle$ are the eigenstates of the bulk Hamiltonian $\hat{\mathcal{H}}_{k}$ and the integral is computed over the first Brillouin zone. For $t<t'$, the Zak phase is $\mathcal{Z}_{1,2}=\pi$ and the system is in the topological phase, while for $t>t'$, the Zak phase is $\mathcal{Z}_{1,2}=0$ and the system is in the trivial phase [see Fig.~\ref{FigTopologicalPhases}(a) and (b)]. As predicted by the bulk-boundary correspondence, the system with open boundary conditions presents two edge states in the topological phase that are not present in the trivial phase, shown in the next section. 

One might try to compute the Zak phase using the eigenstates of the non-rotated Hamiltonian, Eq.~(\ref{EqHamiltonianKspace}), which presents four energy bands and three energy gaps. In that case, the presence of edge states at each gap would be given by the sum of the Zak phases of all the bands below that gap. However, those results do not fulfill the bulk-boundary correspondence, as they do not correctly predict the presence of edge states. This is due to the fact that the Hamiltonian presents a unitary symmetry that exchanges the circulations $+\leftrightarrow -$ at each site. As a result, the Hamiltonian can be block diagonalized such that each pair of bands arising from each symmetry sector have independent Zak phases associated to them.

Figure \ref{FigTopologicalPhases} shows the phases of the symmetric, $\hat{\mathcal{H}}_s$, and antisymmetric, $\hat{\mathcal{H}}_a$, chains as a function of the separation distances $d$ and $d'$ for $\rho_0=5\sigma$. The color represents the ratios (a) $t_s/t_s'$ and (b) $t_a/t_a'$, and the solid black lines separate the trivial (above) and topological (below) phases. We define the nearly dimerized regimes of the trivial and topological phases by their lower boundaries at $t/t'=10$ (dotted blue line) and upper boundaries at $t/t'=0.1$ (dashed yellow line), respectively. In the nearly dimerized regime, the SSH model is well approximated by a set of decoupled dimers that correspond to the dimerized limit. Both chains are in the topological phase for $d>d'$, and in the trivial phase for $d<d'$, but the ratio $t/t'$ varies for the symmetric and antisymmetric chains, as it is determined by the dependence of the couplings $t_a,t_s$ on $J_2$ and $J_3$, Eq.~(\ref{EqCouplingsSSH}). Thus, it is subject to their constraints, namely: (i) the couplings decay with the distance $d$, (ii) $J_3>J_2$ with $J_3\approx J_2$ for $d\gg 1$, (iii) $J_2,J_3>0$, valid for the considered range of distances.

\begin{figure}[t]
	\includegraphics[width=1\columnwidth]{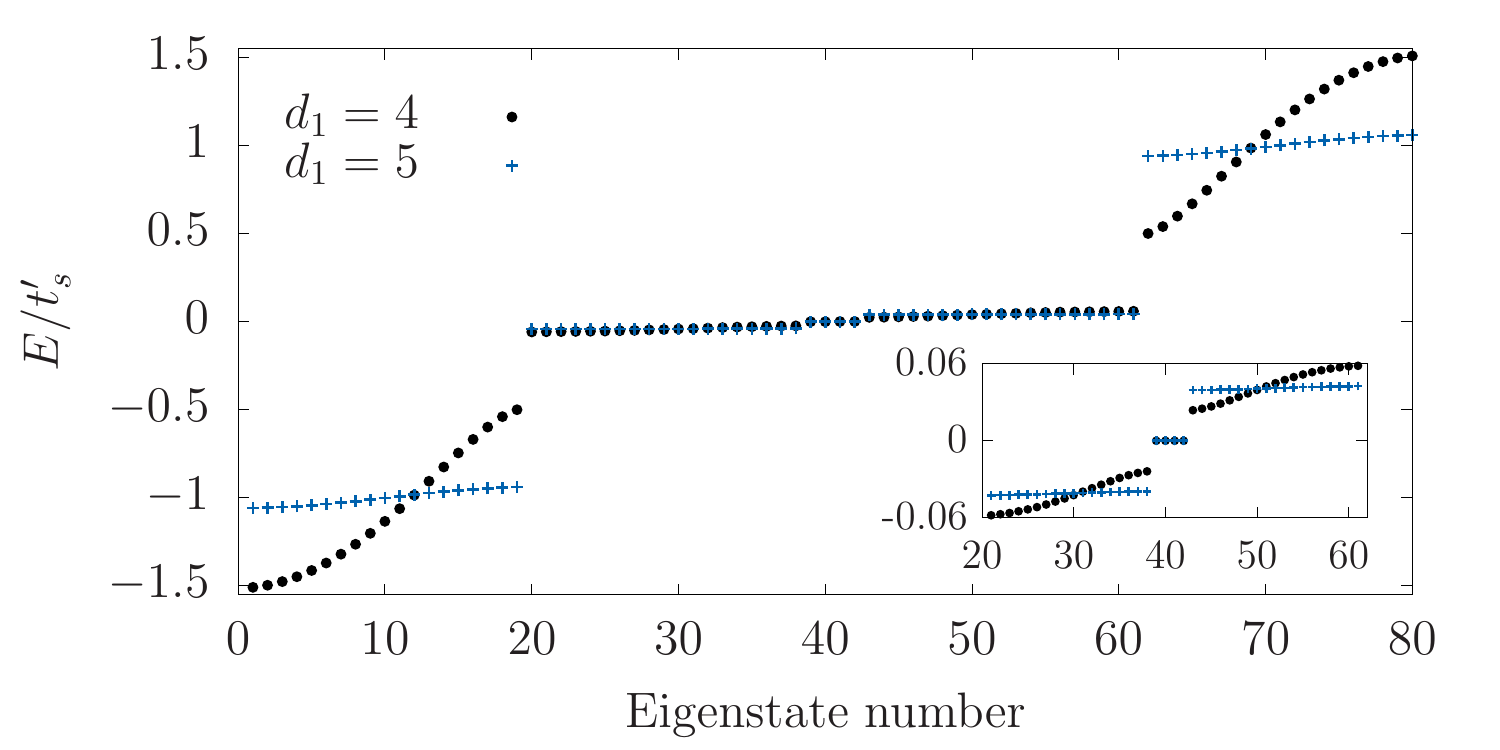}
	\caption{Single particle energy spectrum for a chain of $N_c=20$ unit cells with $\rho_0=5\sigma$, distances $d'=3.6\sigma$, $d=4\sigma$ (black dots) and $d=5\sigma$ (blue crosses). The inset shows the inner bands given by $\hat{\mathcal{H}}_a$ and the zero-energy edge states. }
	\label{FigSpectrumSingleParticle}
\end{figure}

Decoupled SSH chains also emerge in other physical platforms, such as the polariton micropillar system supporting the excited photonic modes $p_x$ and $p_y$ studied in \cite{St-Jean2017}. The zigzag configuration of the micropillar structure gives rise to two decoupled SSH chains corresponding to the $p_x$ and $p_y$ modes, with a glide reflection symmetry between the chains \cite{Zhang2017}, which are in opposite topological phases due to the geometry of the structure. In contrast, here both SSH chains are in the same topological phase for any pair of distances $d$ and $d'$.

\begin{figure*}[t]
	\includegraphics[width=1\columnwidth,trim=0cm 0cm 0cm 0cm,clip]{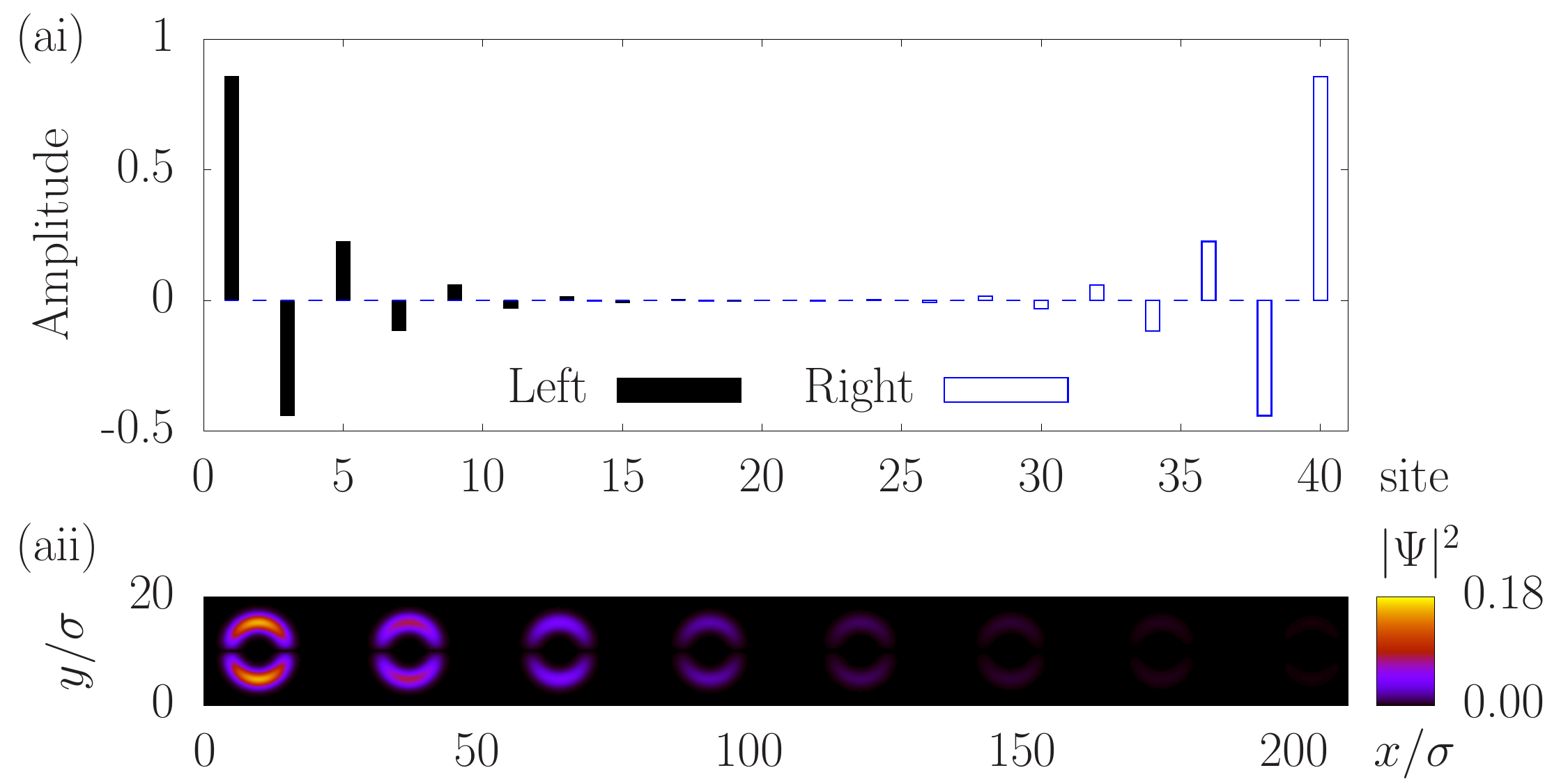}
	\includegraphics[width=1\columnwidth,trim=0cm 0cm 0cm 0cm,clip]{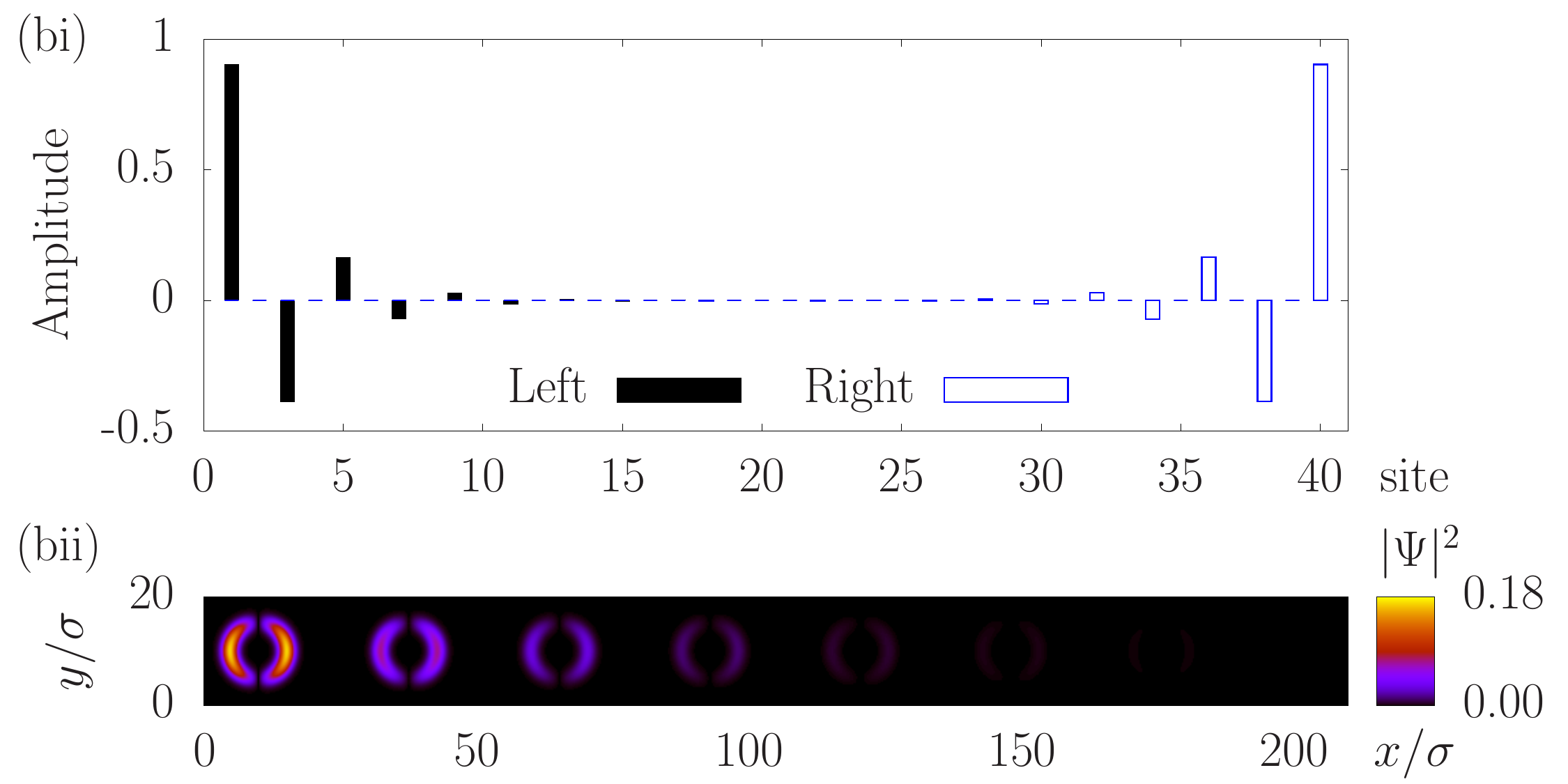}
	\includegraphics[width=1\columnwidth,trim=0cm 0cm 0cm 0cm,clip]{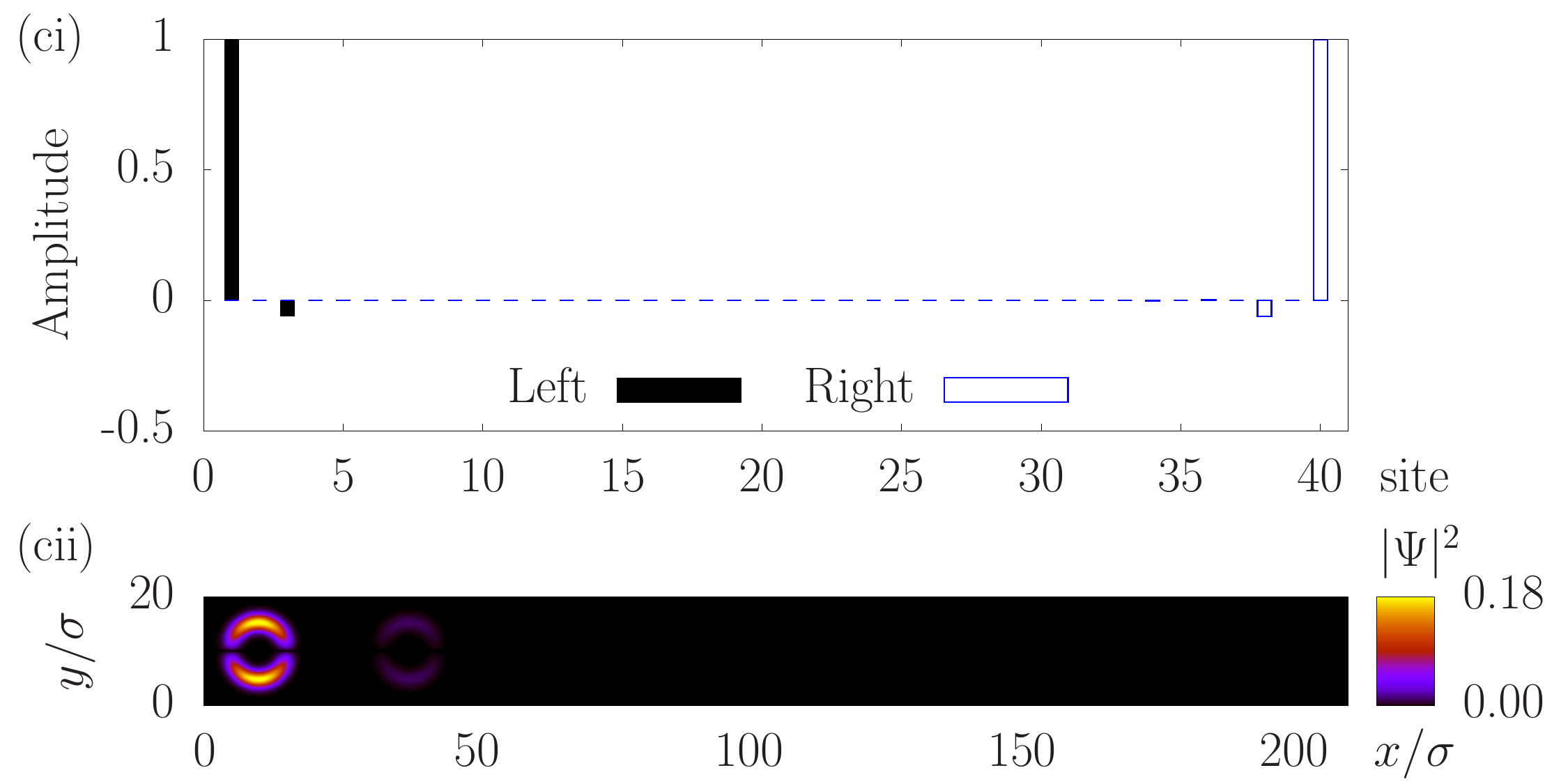}
	\includegraphics[width=1\columnwidth,trim=0cm 0cm 0cm 0cm,clip]{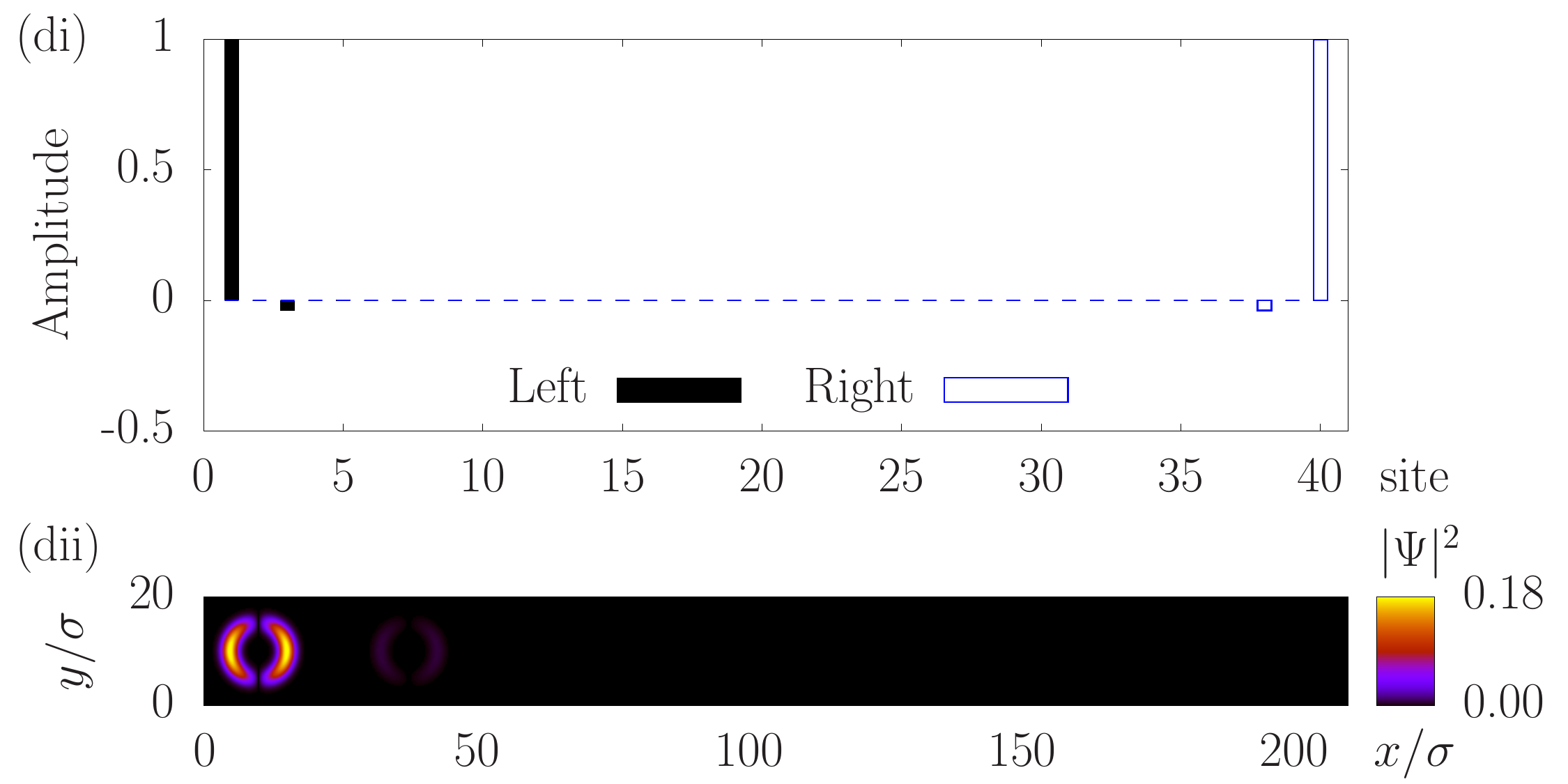}	
	\caption{Edge states of (a),(c) $\hat{\mathcal{H}}_s$ and (b),(d) $\hat{\mathcal{H}}_a$. The distance is $d=4\sigma$, (a) and (b), and $d=5\sigma$, (c) and (d), with $d'=3.6\sigma$, $\rho_0=5\sigma$, and $N_c=20$  for all cases. (i) Amplitude of the left (filled black bars) and right (empty blue bars) edge states of the symmetric and antisymmetric states (\ref{EqBasisChange}) at each site. (ii) Real space density, $|\Psi|^2$ of the first $15$ sites of the left edge state. The edge states arise as the symmetric and antisymmetric superpositions of their hybridised counterparts.
	} \label{FigEdgeStatesSingleParticle}
\end{figure*}

\subsection{Exact diagonalization results}\label{SecSingleExactDiagonalization}
We consider a lattice of rings of $N_c=20$ unit cells with $\rho_0=5\sigma$, $d'=3.6\sigma$ and the two cases $d=4\sigma$ and $5\sigma$. Figure \ref{FigSpectrumSingleParticle} shows the energy spectrum of the system for $d=4\sigma$ (black dots) and $d=5\sigma$ (blue crosses). The outer bands correspond to the symmetric chain $\hat{\mathcal{H}}_s$ [$\epsilon_{3,4}$ in (\ref{EqEnergyBands})], while the inner bands correspond to the antisymmetric chain $\hat{\mathcal{H}}_a$ [$\epsilon_{1,2}$ in (\ref{EqEnergyBands})]. This correspondence can be deduced from the values of $t_s$ and $t_a$ [see Eq.~(\ref{EqCouplingsSSH}) and Fig.~\ref{FigCouplingsVsDistance}], as they fulfill $t_s>t_a$. Additionally, the condition $t_s>t_a$ also makes the outer bands more dispersive than the inner bands for both distances $d$ [see $\epsilon_{1,2}$ and $\epsilon_{3,4}$ in Eq.~(\ref{EqEnergyBands})]. For both cases, the two chains are in the topological phase, which leads to the presence of four edge states, two for each chain. The dispersion of both models is reduced considerably for $d=5\sigma$ compared to $d=4\sigma$, as both chains enter the nearly dimerized regime.

Figure \ref{FigEdgeStatesSingleParticle} shows (i) the amplitude of the left and right edge states of $\hat{\mathcal{H}}_s$ and $\hat{\mathcal{H}}_a$, and (ii) the real space densities $|\Psi|^2$ of the left edge state, taking $d'=3.6\sigma$, $\rho_0=5\sigma$, and $N_c=20$ for all cases. Figures (a) and (c) correspond to the symmetric chain for $d=4\sigma$ and $d=5\sigma$, respectively. Figures (b) and (d) correspond to the antisymmetric chain for $d=4\sigma$ and $d=5\sigma$, respectively. In subfigures (i), each bar represents the amplitude of the basis states in Eq.~(\ref{EqBasisChange}) for each site. In all cases, the edge states only populate one sublattice, either the $A$, or the $B$ sites, with the population decaying exponentially from the edge. They are obtained as the symmetric and antisymmetric superpositions of their hybridised counterparts. The edge states for $d=5\sigma$, show an almost complete localization of the population in the edge site as both $\hat{\mathcal{H}}_s$ and $\hat{\mathcal{H}}_a$ are within the nearly dimerized regime. However, for $d=4\sigma$ all edge states penetrate considerably into the bulk. In subfigures (ii) that show the real space density plots, one can see the difference between the edge states of $\hat{\mathcal{H}}_s$ and $\hat{\mathcal{H}}_a$ in the orientation of the nodes that appear in the density $|\Psi|^2$. These nodes appear due to the superposition of the positive and negative circulations of the mode with orbital angular momentum $l=1$, Eq.~(\ref{EqBasisChange}).

\section{Two-particle}\label{SecTwoParticle} 
\begin{figure*}[t]
	\includegraphics[width=2\columnwidth]{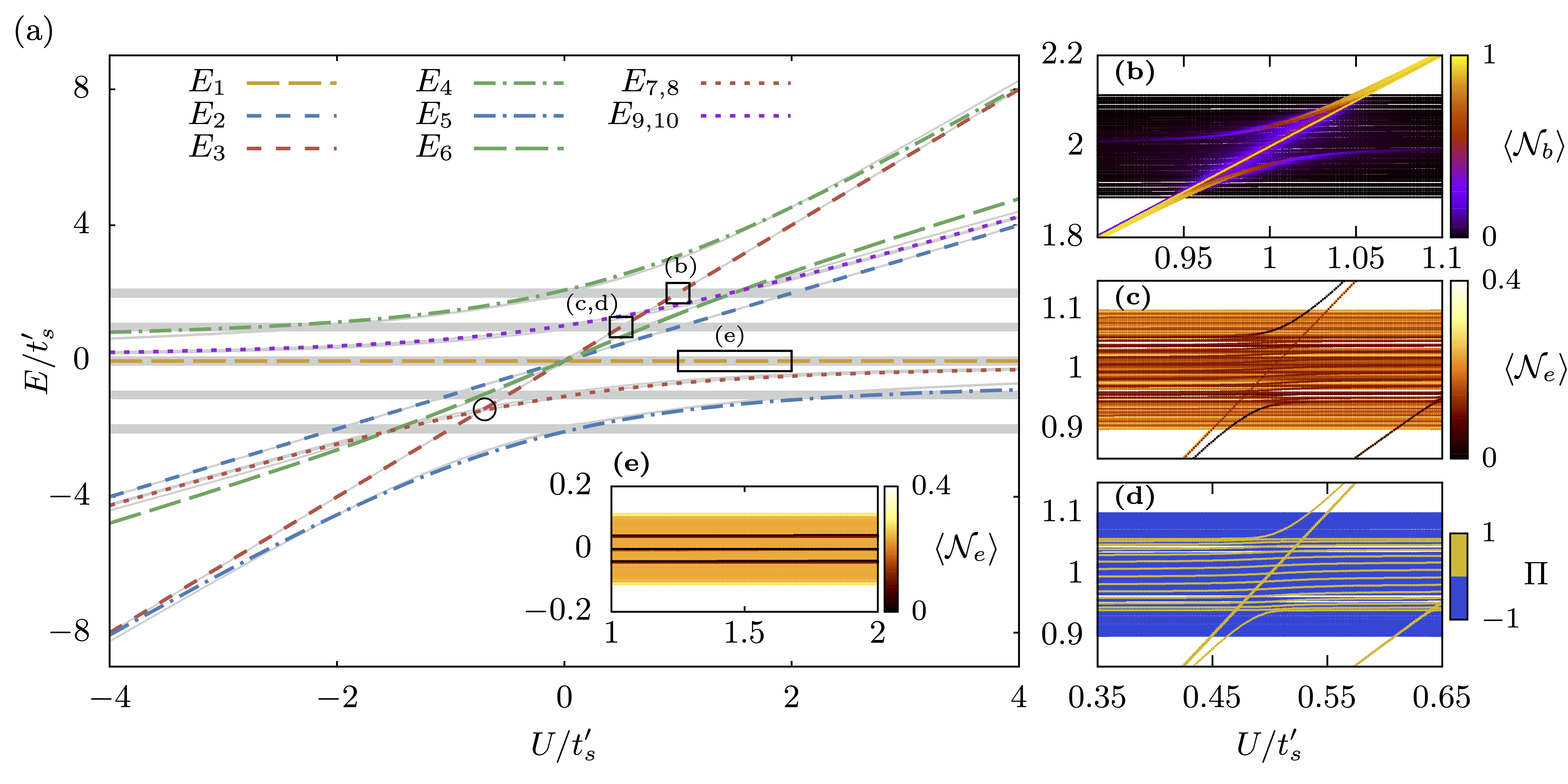}
	\caption{(a) Two-particle energy spectrum for a chain of $N_c=15$ unit cells with $\rho_0=5\sigma$ and distances $d'=3.6\sigma$, $d=5\sigma$ obtained through exact diagonalization (gray lines) and eigenvalues of the strong-link Hamiltonian Eq.~(\ref{EqStrongLinkHamiltonian}) for $J_3'=J$ in color. (b)-(e): Numerical results for various sections of the spectrum with the color indicating (b) the expectation value of the bound state population $\mathcal{N}_b$; (c) and (e), the expectation value of the average distance to the nearest edge $\mathcal{N}_e$, and (d) parity $\Pi$. The circle indicates a crossing between different doublon bands and the rectangles indicate the sections of the spectrum depicted in (b)-(e).}
	\label{FigSpectrumVsU}
\end{figure*}

In this section we investigate the role of on-site bosonic interactions by considering the simplest possible interacting case, a two-boson system. The total Hamiltonian is $\hat{\mathcal{H}}_{l=1}=\hat{\mathcal{H}}_{l=1}^{0}+\hat{\mathcal{H}}_{l=1}^{\mathrm{int}}$, where the independent-particle term $\hat{\mathcal{H}}_{l=1}^{0}$ [interaction term $\hat{\mathcal{H}}_{l=1}^{\mathrm{int}}$] is given in Eq.~(\ref{EqKineticHamiltonianHubbard})  [Eq.~(\ref{EqInteractionHamiltonianHubbard})]. Figure \ref{FigSpectrumVsU} shows the two-particle energy spectrum in gray lines as a function of the interaction strength to tunneling ratio $U/t^\prime_s$ for a chain of $N_c=15$ unit cells with $\rho_0=5\sigma$ and the distances $d=5\sigma$ and $d'=3.6\sigma$. At zero interaction $U/t^\prime_s=0$, the spectrum presents five scattering continua, which correspond to the different two-particle combinations that occupy the different energy bands of the single-particle spectrum. As the particles occupy different sites, the energy bandwidth of these bands stays constant for any value of $U$. For a non-zero interaction strength, nine additional bands can be distinguished, for which the energy depends on the interaction strength $U$. On top of these bands we plot in color the eigenvalues of the strong-link Hamiltonian discussed in the next section. Those bands have contributions of basis states where two bosons occupy the same site, forming a bound state referred as doublon, that leads to a nonzero interaction energy \cite{Hubbard1963,Mattis1986,Winkler2006,Creffield2004,Creffield2010}.

In order to characterize the doublon bands, we analyze two regimes: (i) For the trivial dimerized limit of the original lattice, where $d'\gg d$ and $J_2^\prime,J_3^\prime\simeq0$, we derive a strong-link model that describes the doublon bands at any interaction strength $U$, and (ii) For the limit of strong interactions, $U\gg J_2^{(\prime)},J_3^{(\prime)}$, we use perturbation theory to describe the effective subspaces that appear as a result of introducing the couplings $J_2^{(\prime)},J_3^{(\prime)}$ as a perturbation.

\subsubsection{Strong link Hamiltonian}\label{SecStrongLink}
In this section we consider the dimerized limit of the real space SSH lattice. First we consider the trivial dimerization, for which $d'\gg d$, such that the corresponding couplings can be neglected, $J_2^\prime,J_3^\prime\ll J_2,J_3$. Thus, the symmetric $\hat{\mathcal{H}}_s$ and antisymmetric $\hat{\mathcal{H}}_a$ SSH lattices are also in the dimerized limit. In this regime, the doublon bands can be described by the reduced two-particle Hamiltonian of a single strong link of a unit cell $m$. This approach was used in \cite{DiLiberto2016} to analyze the doublon bands of the two-particle conventional SSH model. The basis states of this two-particle Hamiltonian are the ten two-particle combinations of the sites $A$ and $B$ in a unit cell $m$, namely: 
\begin{widetext}
	\begin{equation}\label{EqBasisStrongLink}
		\begin{aligned}
			\big\{\left|A_{m}^+ A_{m}^+\right\rangle,
			\left|A_{m}^+ A_{m}^-\right\rangle,
			\left|A_{m}^+ B_{m}^+\right\rangle,
			\left|A_{m}^+ B_{m}^-\right\rangle,
			\left|A_{m}^- A_{m}^-\right\rangle,
			\left|A_{m}^- B_{m}^+\right\rangle,
			\left|A_{m}^- B_{m}^-\right\rangle,
			\left|B_{m}^+ B_{m}^+\right\rangle,
			\left|B_{m}^+ B_{m}^-\right\rangle,
			\left|B_{m}^- B_{m}^-\right\rangle\big\}.
		\end{aligned}
	\end{equation}
In this basis, the strong-link Hamiltonian reads

\begin{equation}\label{EqStrongLinkHamiltonian}
	\hat{\mathcal{H}}_{SL}=\begin{pmatrix}
		U & 0 &\sqrt{2}J_2 &\sqrt{2}J_3&0&0&0&0&0&0\\
		0&2U&J_3&J_2&0&J_2&J_3&0&0&0\\
		\sqrt{2}J_2&J_3&0&0&0&0&0&\sqrt{2}J_2&J_3&0\\
		\sqrt{2}J_3&J_2&0&0&0&0&0&0&J_2&\sqrt{2}J_3\\
		0&0&0&0&U&\sqrt{2}J_3&\sqrt{2}J_2&0&0&0\\
		0&J_2&0&0&\sqrt{2}J_3&0&0&\sqrt{2}J_3&J_2&0\\
		0&J_3&0&0&\sqrt{2}J_2&0&0&0&J_3&\sqrt{2}J_2\\
		0&0&\sqrt{2}J_2&0&0&\sqrt{2}J_3&0&U&0&0\\
		0&0&J_3&J_2&0&J_2&J_3&0&2U&0\\
		0&0&0&\sqrt{2}J_3&0&0&\sqrt{2}J_2&0&0&U
	\end{pmatrix}.
\end{equation}

\begin{table*}[t]
	\begin{center}
		\begin{tabular}{c c c c c c c c c c c c c}
			\hline\hline\rule{0pt}{12pt}
			$v_n$ &$E_{n,|U|\gg J}$&$\Pi$&
			$\left|A_{m}^+ A_{m}^+\right\rangle$&
			$\left|A_{m}^+ A_{m}^-\right\rangle$&
			$\left|A_{m}^+ B_{m}^+\right\rangle$&
			$\left|A_{m}^+ B_{m}^-\right\rangle$&
			$\left|A_{m}^- A_{m}^-\right\rangle$&
			$\left|A_{m}^- B_{m}^+\right\rangle$&
			$\left|A_{m}^- B_{m}^-\right\rangle$&
			$\left|B_{m}^+ B_{m}^+\right\rangle$&
			$\left|B_{m}^+ B_{m}^-\right\rangle$&
			$\left|B_{m}^- B_{m}^-\right\rangle$\\[1mm]\hline
			$v_1$&$0$&$+$&$0$&$0$&$\frac{1}{2}$&$-\frac{1}{2}$&$0$&$-\frac{1}{2}$&$\frac{1}{2}$&$0$&$0$&$0$\\
			$v_2$&$U$&$+$&$-\frac{1}{2}$&$0$&$0$&$0$&$-\frac{1}{2}$&$0$&$0$&$\frac{1}{2}$&$0$&$\frac{1}{2}$\\
			$v_3$&$2U$&$+$&$0$&$-\frac{1}{\sqrt{2}}$&$0$&$0$&$0$&$0$&$0$&$0$&$\frac{1}{\sqrt{2}}$&$0$\\
			$v_4$&$\{2U,0\}$&$+$&$0$&$\{\frac{1}{\sqrt{2}},0\}$&$\{0,\frac{1}{2}\}$&$\{0,\frac{1}{2}\}$&$0$&$\{0,\frac{1}{2}\}$&$\{0,\frac{1}{2}\}$&$0$&$\{\frac{1}{\sqrt{2}},0\}$&$0$\\
			$v_5$&$\{0,2U\}$&$+$&$0$&$\{0,\frac{1}{\sqrt{2}}\}$&$\{-\frac{1}{2},0\}$&$\{-\frac{1}{2},0\}$&$0$&$\{-\frac{1}{2},0\}$&$\{-\frac{1}{2},0\}$&$0$&$\{0,\frac{1}{\sqrt{2}}\}$&$0$\\
			$v_6$&$U$&$+$&$\frac{1}{2}$&$0$&$0$&$0$&$\frac{1}{2}$&$0$&$0$&$\frac{1}{2}$&$0$&$\frac{1}{2}$\\
			$v_7$&$\{0,U\}$&$-$&$\{0,\frac{1}{2}\}$&$0$&$0$&$\{-\frac{1}{\sqrt{2}},0\}$&$\{0,-\frac{1}{2}\}$&$\{\frac{1}{\sqrt{2}},0\}$&$0$&$\{0,-\frac{1}{2}\}$&$0$&$\{0,\frac{1}{2}\}$\\
			$v_8$&$\{0,U\}$&$-$&$\{0,\frac{1}{2}\}$&$0$&$\{-\frac{1}{\sqrt{2}},0\}$&$0$&$\{0,-\frac{1}{2}\}$&$0$&$\{\frac{1}{\sqrt{2}},0\}$&$\{0,\frac{1}{2}\}$&$0$&$\{0,-\frac{1}{2}\}$\\
			$v_9$&$\{U,0\}$&$-$&$\{\frac{1}{2},0\}$&$0$&$0$&$\{0,\frac{1}{\sqrt{2}}\}$&$\{-\frac{1}{2},0\}$&$\{0,-\frac{1}{\sqrt{2}}\}$&$0$&$\{-\frac{1}{2},0\}$&$0$&$\{\frac{1}{2},0\}$\\
			$v_{10}$&$\{U,0\}$&$-$&$\{\frac{1}{2},0\}$&$0$&$\{0,\frac{1}{\sqrt{2}}\}$&$0$&$\{-\frac{1}{2},0\}$&$0$&$\{0,-\frac{1}{\sqrt{2}}\}$&$\{\frac{1}{2},0\}$&$0$&$\{-\frac{1}{2},0\}$
		\end{tabular}
	\end{center}
	\caption{Normalized eigenvectors $v_n$ and eigenvalues $E_n$ of the strong-link Hamiltonian $\hat{\mathcal{H}}_{SL}$ in the limit of large distances $d \gg 1$, and the regime of strong interactions $|U|\gg J$. Two cases are considered, $\{U >0,U<0\}$, and a single number is given when it is the same for both cases. Note that the eigenvectors and eigenvalues $n=1,2,3$ are independent of the ratio $|U|/J$. The third column indicates the parity $\Pi$ of the eigenstate with respect to the exchange circulation symmetry $+\leftrightarrow -$ in each site.}
	\label{TableEigenvectorsLimit}
\end{table*}

To find the eigenvalues and eigenvectors of $\hat{\mathcal{H}}_{SL}$, we consider the limit of $d \gg 1$, for which $J_2=J_3=J$. In this limit, the eigenvalues are 
\begin{equation}\begin{aligned}\label{EqEigenvaluesStrongLink}
		E_1=&0, \qquad\quad E_2=U, \qquad\quad E_3=2U \qquad E_{7,8}=\dfrac{U-\sqrt{16J^2+U^2}}{2}, \quad
		E_{9,10}=\dfrac{U+\sqrt{16J^2+U^2}}{2},\\
		E_{4+q} =&\left[\hspace{-0.3mm}\cos\hspace{-0.5mm}\left(\frac{1}{3}\arccos\left[\hspace{-0.5mm}-\frac{U}{4\sqrt{2} J}\left(\frac{\frac{U^2}{8 J^2}+2}{3}\right)^{\hspace{-1mm}\frac{-3}{2}}\,\right]\hspace{-1mm}+\frac{2\pi q}{3}\right)\right.\left.\times 2\sqrt{\frac{\frac{U^2}{8 J^2}+2}{3}}+\frac{U}{2\sqrt{2} J}\,\,\right] 2\sqrt{2} J,  \,\, \mbox{for } q=0,1,2.
	\end{aligned}
\end{equation}
\end{widetext}
Table \ref{TableEigenvectorsLimit} presents the eigenvectors in the limit $|U|\gg J$ while the general expressions can be found in Table~\ref{TableEigenvectors} of the Appendix. The  strong-link model for the topological dimerization is analogous to the trivial dimerization model and it can be obtained by simply replacing the indices $m$ of each $A$ site in the basis states (\ref{EqBasisStrongLink}) by $m+1$. This yields the same eigenvectors and eigenvalues given in Eq.~(\ref{EqEigenvaluesStrongLink}) and Tables \ref{TableEigenvectorsLimit} and \ref{TableEigenvectors} in the limit $d \gg 1$. 

Figure \ref{FigSpectrumVsU}(a) shows the eigenvalues (in color) on top of the exact diagonalization results (in gray) as a function of the ratio $U/t_s^\prime$. The exact diagonalization results correspond to the distances $d=5\sigma$ and $d'=3.6\sigma$, which determine the values of the couplings $J_2^{(\prime)},J_3^{(\prime)}$ (see Fig.~\ref{FigCouplingsVsDistance}). To compare the analytical and numerical results, we fix the coupling $J$ of $\hat{\mathcal{H}}_{SL}$ as the largest numerical coupling, $J_3^\prime$, which corresponds to the topological dimerization of the strong-link model. Thus, we consider the eigenvalues of $\hat{\mathcal{H}}_{SL}$ in the limit $d \gg 1$ (for which $J_2=J_3=J$), Eq.~(\ref{EqEigenvaluesStrongLink}), which are common to both dimerizations. The analytically obtained eigenvalues accurately predict the overall energy dependence of the doublon bands obtained numerically. 

Table \ref{TableEigenvectorsLimit} gives the eigenvectors $v_n$ of the strong-link Hamiltonian Eq.~(\ref{EqStrongLinkHamiltonian}) in the regime of strong interactions $|U|\gg J$ for two cases, $\{U >0,U<0\}$, and specifies the corresponding eigenvalues $E_n$ for the two regimes. The eigenvectors and eigenvalues $n=\{1-3\}$ do not depend on the coupling $J$ and thus they do not change when increasing $|U|/J$. In both regimes, the eigenvalues form three groups, with energies $0$, $U$ and $2U$. This tendency can be observed even for relatively small ratios $|U|/t_s^\prime$ in Fig.~\ref{FigSpectrumVsU}, where the energy difference between the different groups of doublon bands diminishes for increasing values of $|U|/t_s^\prime$. The eigenvectors in Table \ref{TableEigenvectorsLimit} clarify the differences between these three groups. The eigenvectors with energy zero are a superposition of states where the two particles populate different sites, $|A_m^{\alpha},B_m^{\alpha^\prime}\rangle$. As there are four possible states, there are four bands with energy zero. In these states the two bosons do not interact, and their energy, which is given by the interaction Hamiltonian term (\ref{EqInteractionHamiltonianHubbard}), becomes zero. The group with energy $U$ is composed of eigenstates that are a superposition of states with two bosons in the same site and the same circulation, $|j_m^{\alpha},j_m^{\alpha}\rangle$, which also results in four possible states and four energy bands. Finally, the group with energy $2U$ is formed of states where the bosons occupy opposite circulations in the same site $|j_m^{+},j_m^-\rangle$. In this case, there are only two available states and this results in only two energy bands.

\begin{figure}[t]
	\includegraphics[width=0.8\columnwidth]{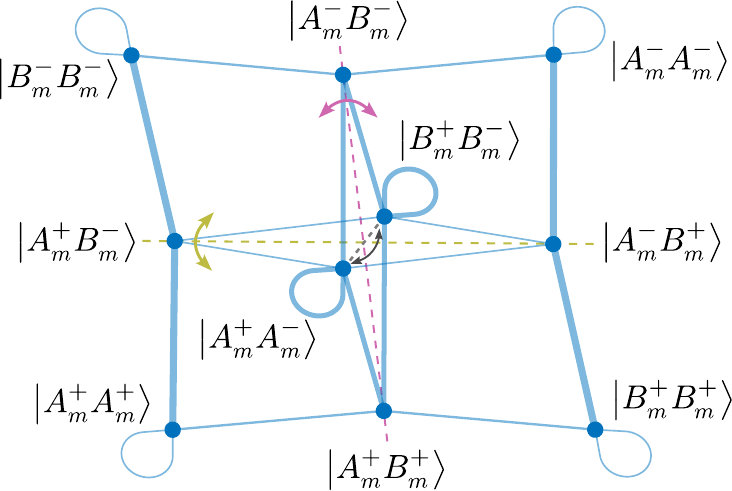}
	\caption{Adjacency graph of the strong-link Hamiltonian $\hat{\mathcal{H}}_{SL}$ Eq.~(\ref{EqStrongLinkHamiltonian}). Each vertex represents one of the basis states in Eq.~(\ref{EqBasisStrongLink}) and the edges represent the off-diagonal (solid blue lines) and diagonal (solid blue loops) matrix elements of $\hat{\mathcal{H}}_{SL}$. The dashed lines indicate the reflection symmetries with respect to the $oxz$ (yellow), $oyz$ (pink), and $oxy$ planes, where the states $|B_m^+B_m^-\rangle$ and $|A_m^+A_m^-\rangle$ are located in the out-of-plane axis $z$ (gray).}
	\label{FigGraphStrongLink}
\end{figure}

Figure \ref{FigGraphStrongLink} represents the adjacency graph of the strong-link Hamiltonian $\hat{\mathcal{H}}_{SL}$ and its symmetries. Each vertex represents one of the basis states in Eq.~(\ref{EqBasisStrongLink}) and the edges represent the off-diagonal (solid blue lines) and diagonal (solid blue loops) matrix elements of $\hat{\mathcal{H}}_{SL}$. The symmetries of the graph become explicit by locating the states $|B_m^+B_m^-\rangle$ and $|A_m^+A_m^-\rangle$ in the out-of-plane $z$ axis, which is perpendicular to the plane where all the other basis states lie. Then, the graph explicitly exhibits three symmetries: (i) a reflection symmetry with respect to the $oxz$ (yellow) plane that leaves the states $|B_m^+B_m^-\rangle$, $|A_m^+A_m^-\rangle$, $|A_m^+B_m^-\rangle$, and $|A_m^-B_m^+\rangle$ invariant; (ii) a reflection symmetry with respect to the $oyz$ (pink) plane that leaves the states $|B_m^+B_m^-\rangle$, $|A_m^+A_m^-\rangle$, $|A_m^-B_m^-\rangle$, and $|A_m^+B_m^+\rangle$ invariant; and (iii) a reflection symmetry respect to $oxy$ plane that only permutes the states $|B_m^+B_m^-\rangle$ and $|A_m^+A_m^-\rangle$ located in the out-of-plane axis $z$ (gray). A single strong-link is symmetric with respect to the exchange of the sites $A$ and $B$, regardless of the number of particles. This symmetry is reflected in the two-particle graph as the combined application of the symmetries (ii) and (iii). However, this symmetry disappears when we move away from the dimerized limit by introducing weak couplings between the different strong links. In contrast, there is a symmetry that originates in the total Hamiltonian [Eqs.~(\ref{EqKineticHamiltonianHubbard}) and (\ref{EqInteractionHamiltonianHubbard})], the exchange of circulations $+\leftrightarrow -$ in each site, which is inherited by $\hat{\mathcal{H}}_{SL}$. This symmetry can be obtained by applying the symmetries (i) and (ii) in the two-particle graph. Both the eigenstates of $\hat{\mathcal{H}}_{l=1}$ and $\hat{\mathcal{H}}_{SL}$ have well-defined parities $\Pi$ with respect to the exchange of circulations in each site, and we indicate the latter in Table \ref{TableEigenvectorsLimit}.

The spectrum in Fig.~\ref{FigSpectrumVsU} shows a large number of intersections and avoided crossings between the doublon bands and the scattering continua, as well as  between different doublon bands at $U/t_s^\prime \simeq\pm0.7$ [black circle in Fig.~\ref{FigSpectrumVsU}(a)]. The presence of avoided crossings can be understood in terms of the circulation exchange symmetry $+\leftrightarrow -$ discussed above. They can appear for doublon states and extended states of the same parity that converge in energy. In contrast, states of opposite parities belong to different symmetry sectors and are therefore completely decoupled. Below we discuss two examples of avoided crossings.

Fig. \ref{FigSpectrumVsU}(b) shows the avoided crossing of the state $v_3$ with the upper band of extended states. The color represents the expectation value of the bound state population in each strong link
$\mathcal{N}_b=\sum_{m=1}^{N_c-1}\hat{n}_m^b$,
where
\begin{equation}
	\begin{aligned}
		\hat{n}_m^b=& \hat{n}_{A_{m+1}}^+\hat{n}_{A_{m+1}}^++\hat{n}_{A_{m+1}}^+\hat{n}_{A_{m+1}}^-+\hat{n}_{A_{m+1}}^+\hat{n}_{B_{m}}^+\\
		&+\hat{n}_{A_{m+1}}^+\hat{n}_{B_{m}}^-+\hat{n}_{A_{m+1}}^-\hat{n}_{A_{m+1}}^-	+\hat{n}_{A_{m+1}}^-\hat{n}_{B_{m}}^+\\
		&+\hat{n}_{A_{m+1}}^-\hat{n}_{B_{m}}^-+\hat{n}_{B_{m}}^+\hat{n}_{B_{m}}^++\hat{n}_{B_{m}}^+\hat{n}_{B_{m}}^-+\hat{n}_{B_{m}}^-\hat{n}_{B_{m}}^-.
	\end{aligned}
\end{equation}
The number operators account for the ten state combinations that form the strong-link Hamiltonian basis in the topological dimerization. This avoided crossing occurs due to the strong hybridization between the state $v_3$, which has positive parity, and extended states of the upper band, where all states also have positive parity. 

In the avoided crossing shown in Figs.~\ref{FigSpectrumVsU}(c) and (d), the strong resonance involves an edge bound state instead of the states of the band associated to the state $v_3$. The color in Fig.~\ref{FigSpectrumVsU}(c) gives the expectation value of $\mathcal{N}_e=\sum_{o=1}^{2N_c}\sum_{\alpha=\pm} \min((o-1)/2N_c,1-o/2N_c)\hat{n}_{o,\alpha}/2$ where $\hat{n}_{o,\alpha}$ is the number operator at site $n$, so that it represents the average normalized distance to the nearest edge. There are two states below and above the doublon bands with high edge localization (dark lines), which indicates the presence of edge bound states associated to that doublon band. For the same avoided crossing, the color in Fig.~\ref{FigSpectrumVsU}(d) indicates the parity $\Pi$ of each eigenstate. Embedded within the band of extended states, there is an inner band of extended states with a lower value of $\langle \mathcal{N}_e\rangle$, which correspond to those states where a particle occupies the upper single-particle band of the symmetric SSH spectrum and the other particle occupies an edge state  \cite{Marques2017}. Of these states, only some have a positive parity that allows for the avoided crossing to occur. Such embedded states are only present in the topological dimerization of the original chain, which possesses four topologically protected single-particle edge states. They are also present in the zero energy band [dark horizontal lines in Fig.~\ref{FigSpectrumVsU}(e)], in which either the two particles occupy an edge state, at exactly zero energy, or one occupies one edge state while the other occupies the single-particle bands of the antisymmetric SSH lattice.

To obtain the strong-link Hamiltonian we consider that each strong link is completely decoupled from the adjacent ones, such that every dimer yields exactly the same eigenvalues. This model is strictly valid in the dimerized limit, where the either the inter- or intra-cell couplings are exactly zero. As a result of the weak coupling between strong links, the doublon bands computed through exact diagonalization are not exactly degenerate, but present some dispersion [see Fig.~\ref{FigSpectrumVsU}(b)]. Additionally, the doublon bands obtained numerically can present edge bound states above or below the energy of the corresponding doublon band [dark lines in  Fig.~\ref{FigSpectrumVsU}(c)]. They can be identified with Tamm-Shockley states as they are a result of the renormalized couplings at the edge sites that arise when one introduces the weak couplings as a perturbation \cite{DiLiberto2016,Bello2016,Gorlach2017,Salerno2018}. In the next section we are going to lift this degeneracy by moving away from the dimerized limit and we will consider the regime of strong interactions to better understand the emergence of the doublon subspaces.

\subsubsection{Strong interactions limit}
Consider the regime of strong interactions, where the interactions dominate over the tunneling processes, $U\gg J_2,J_3,J_2^\prime,J_3^\prime$. The available states for $U\rightarrow\infty$, can be deduced from the three groups of eigenvectors of the strong-link Hamiltonian for strong interactions (see Table \ref{TableEigenvectorsLimit}). The two bosons can either occupy different sites or occupy the same site forming a doublon. The bosons can form two possible bound states: $\mathcal{A}$, where the two particles occupy the same site and the same circulation, $|j_m^{\alpha},j_m^{\alpha}\rangle$, with $j=A,B$ and $\alpha=\pm$, and $\mathcal{B}$, where the two particles occupy the same site and opposite circulations $|j_m^{+},j_m^-\rangle$. These bound states have energies $E_{\mathcal{A}}=U$ and $E_{\mathcal{B}}=2U$, respectively, given by the interaction term (\ref{EqInteractionHamiltonianHubbard}). 

If one introduces the couplings $J_2,J_3,J_2^\prime,J_3^\prime$ as a perturbation, the bound states in adjacent sites become coupled through second-order hopping processes, which creates an effective dispersive subspace for each bound state class. As the bound states are well separated in energy, these subspaces are decoupled and can be analyzed independently. The matrix elements of the effective Hamiltonian for each subspace up to second-order perturbation theory read 
\cite{Bir1974,Tannoudji1992}
\begin{equation}\label{EqEffectiveHamiltonian}
	\begin{aligned}\langle u|\hat{\mathcal{H}}_{\mathrm{eff}}| u^{\prime}\rangle=& E_{u}^{0} \delta_{u u^{\prime}}+\frac{1}{2} \sum_{w}\langle u|\hat{\mathcal{H}}_{l=1}^0| w\rangle\langle w|\hat{\mathcal{H}}_{l=1}^0| u^{\prime}\rangle\\
		&\times\!\!\left[\frac{1}{E_{u}^{0}-E_{w}^{0}}+\frac{1}{E_{u^{\prime}}^{0}-E_{w}^{0}}\right], 
	\end{aligned}
\end{equation}
where $|u\rangle,|u'\rangle$ are the bound-states, $|w\rangle,|w'\rangle$ are the mediating states in each hopping process, and $E^0$ are the unperturbed energies. Note that the first-order corrections are always zero. For $|u\rangle\neq|u'\rangle$, one obtains an effective tunneling term, while for $|u\rangle=|u'\rangle$, one obtains an effective on-site potential.

\begin{figure}[t]
	\includegraphics[width=1\columnwidth]{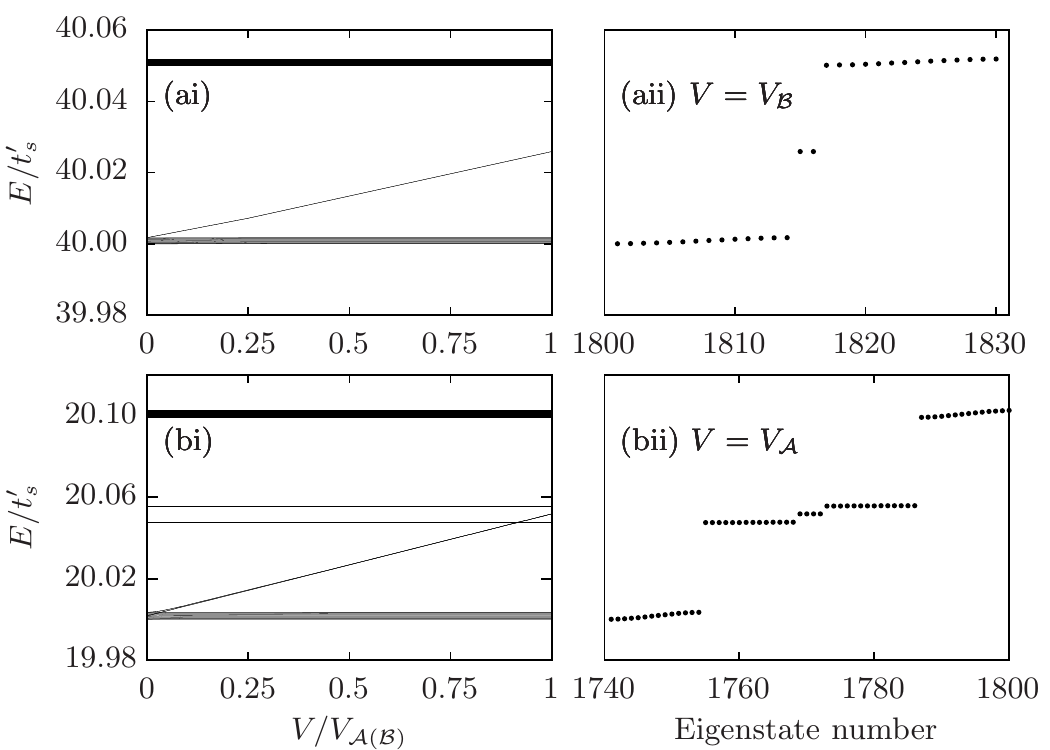}
	\caption{Two-particle energy spectrum for $N_c=15$ unit cells with $\rho_0=5\sigma$, distances $d'=3.6\sigma$, $d=4.5\sigma$, and  $U/t_s^\prime=20$ for the (a) $\mathcal{B}$ and (b) $\mathcal{A}$ subspaces. (i) Energies as a function of the edge on-site potential correction $V$ in units of $V_{\mathcal{A}(\mathcal{B})}$ and (ii) energies for the exactly compensated on-site potential mismatch, (aii) $V=V_\mathcal{B}$ and (bii) $V=V_\mathcal{A}$. }
	\label{FigPotentialCorrection}
\end{figure}

\subsubsection{$\mathcal{B}$ subspace}
Let us start with the $\mathcal{B}$ subspace, which is composed of only one bound state per site, $|j_m^{+},j_m^-\rangle$. The existence of this bound state is due to the inter-circulation interaction term in the interaction Hamiltonian (\ref{EqInteractionHamiltonianHubbard}). Thus, it is a consequence of the ring structure of each site of the lattice \cite{Pelegri2020,Nicolau2022} and cannot appear in a conventional SSH lattice \cite{DiLiberto2016}. When we introduce the couplings $J_2,J_3,J_2^\prime,J_3^\prime$ as a perturbation, the bound states of the $\mathcal{B}$ subspace in adjacent sites become coupled through second-order hopping processes. These yield two effective couplings: an intra-cell coupling between the bound states $|A_m^{+},A_m^-\rangle$ and $|B_m^{+},B_m^-\rangle$, and an inter-cell coupling between $|B_m^{+}B_m^-\rangle$ and $|A_{m+1}^{+},A_{m+1}^-\rangle$. Additionally, each bound state acquires a self-energy term through through second-order processes, yielding effective on-site potentials at each site. Note that in the edge sites there are half the mediating states present in the bulk sites \cite{Bello2016,DiLiberto2016,Marques2017}. Thus, the effective on-site potential in the edge $V_E$ is smaller than the on-site potential in the bulk $V_B$. Using Eq.~(\ref{EqEffectiveHamiltonian}), the resulting effective subspace is an SSH chain with renormalized couplings and a bulk-edge on-site potential mismatch
\begin{equation}\label{EqHamiltonianB}
	\begin{aligned}
		\hat{\mathcal{H}}_\mathcal{B}\!=\!&\hspace{-0.6mm}\left[\dfrac{J_2^2+J_3^2}{U}\sum_{m=1}^{N_c}\hat{a}_m^{\mathcal{B}\dagger}\hat{b}_m^{\mathcal{B}}+\!\dfrac{J_2'^2+J_3'^2}{U}\!\sum_{m=1}^{N_c-1}\!\hat{a}_{m+1}^{\mathcal{B}\dagger}\hat{b}_m^{\mathcal{B}}\right]\hspace{-1.6mm}+\hspace{-0.4mm}\mathrm{H.c.}\\
		&+\dfrac{J_2^2+J_3^2+J_2'^2+J_3'^2}{U}\sum_{m=1}^{N_c}(\hat{a}_m^{\mathcal{B}\dagger}\hat{a}_m^{\mathcal{B}}+\hat{b}_m^{\mathcal{B}\dagger}\hat{b}_m^{\mathcal{B}})\\
		&-\dfrac{J_2^{\prime2}+J_3^{\prime2}}{U}(\hat{a}_1^{\mathcal{B}\dagger}\hat{a}_1^{\mathcal{B}}+\hat{b}_{N_c}^{\mathcal{B}\dagger}\hat{b}_{N_c}^{\mathcal{B}}),
	\end{aligned}
\end{equation}
where the creation and annihilation operators $\hat{a}_m^{\mathcal{B}(\dagger)}$ and $\hat{b}_m^{\mathcal{B}(\dagger)}$ correspond to the bound states of the $\mathcal{B}$ subspace in the $A$ and $B$ sites, respectively. This mismatch can be exactly compensated by introducing an on-site potential $V$ in the edge sites of the real space lattice. Figure \ref{FigPotentialCorrection}(a) shows the energy spectrum of the $\mathcal{B}$ subspace for $N_c=15$ unit cells with $\rho_0=5\sigma$, distances $d'=3.6\sigma$, $d=4.5\sigma$, and  $U/t_s^\prime=20$. In Fig. \ref{FigPotentialCorrection}(ai), the spectrum is represented as a function of the edge potential correction $V$ in units of $V_\mathcal{B}=(J_2'^2+J_3'^2)/2U$, the value of the potential that exactly compensates the mismatch. Figure ~\ref{FigPotentialCorrection}(aii) shows the spectrum for  the potential $V_\mathcal{B}$. For $V=0$, the spectrum presents only two  dispersive bands. The lower band contains two extra states that arise due to the impurity potentials at the edge sites of the SSH chain. When increasing the potential correction $V$, these Tamm-Shockley states depart from the lower chain as their energy grows linearly and become localized at the edge. For $V_\mathcal{B}$, the potential mismatch is compensated exactly, thus restoring the chiral symmetry of the model and yielding two topologically-protected edge states.

\subsubsection{$\mathcal{A}$ subspace}\textit{}
In contrast with the $\mathcal{B}$ subspace, the $\mathcal{A}$ subspace presents two bound states per site instead of only one, $|j_m^{\alpha},j_m^{\alpha}\rangle$ with $\alpha=\pm$. In the strong interactions regime, the bound states in adjacent sites become coupled through second-order hoppings, which yields four effective couplings between the adjacent bound states $|A_m^{\alpha},A_m^{\alpha}\rangle$ and $|B_{m'}^{\alpha'},B_{m'}^{\alpha'}\rangle$, thus forming a Creutz ladder structure \cite{Zurita2020}. In analogy with the $\mathcal{B}$ subspace, each bound state also obtains an effective on-site potential that generates a bulk-edge on-site potential mismatch. Using Eq.~(\ref{EqEffectiveHamiltonian}), the effective model of this subspace reads
\begin{equation}\label{EqEffectiveA}
	\begin{aligned}
		\hat{\mathcal{H}}_\mathcal{A}=
		&\sum_{\alpha=\pm}\Bigg(\sum_{m=1}^{N_c}\Bigg[\dfrac{2J_2^2}{U}\hat{a}_m^{\mathcal{A}_\alpha\dagger}\hat{b}_m^{\mathcal{A}_\alpha}+\dfrac{2J_3^2}{U}\hat{a}_m^{\mathcal{A}_\alpha\dagger}\hat{b}_m^{\mathcal{A}_{-\alpha}}\Bigg]+\mathrm{H.c.}\\
		&+\sum_{m=1}^{N_c-1}\Bigg[\dfrac{2J_2'^2}{U}\hat{a}_{m+1}^{\mathcal{A}_\alpha\dagger}\hat{b}_m^{\mathcal{A}_\alpha}+\dfrac{2J_3'^2}{U}\hat{a}_{m+1}^{\mathcal{A}_\alpha\dagger}\hat{b}_m^{\mathcal{A}_{-\alpha}}\Bigg]+\mathrm{H.c.}\\
		&+2\dfrac{J_2^2+J_3^2+J_2'^2+J_3'^2}{U}\sum_{m=1}^{N_c}\left[\hat{a}_m^{\mathcal{A}_\alpha\dagger}\hat{a}_m^{\mathcal{A}_\alpha}+\hat{b}_m^{\mathcal{A}_\alpha\dagger}\hat{b}_m^{\mathcal{A}_\alpha}\right]\\
		&-2\dfrac{J_2^{\prime2}+J_3^{\prime2}}{U}\left[\hat{a}_1^{\mathcal{A}_\alpha\dagger}\hat{a}_1^{\mathcal{A}_\alpha}+\hat{b}_{N_c}^{\mathcal{A}_\alpha\dagger}\hat{b}_{N_c}^{\mathcal{A}_\alpha}\right]\Bigg),
	\end{aligned}
\end{equation}
where the creation and annihilation operators $\hat{a}_m^{\mathcal{A}_\alpha(\dagger)}$ and $\hat{b}_m^{\mathcal{A}_\alpha(\dagger)}$ correspond to the bound states of the $\mathcal{A}$ subspace with two particles in circulation $\alpha$ in the $A$ and $B$ sites, respectively. The effective model of the $\mathcal{A}$ subspace takes the same form as the original single-particle model in Eq.~(\ref{EqKineticHamiltonianHubbard}), with additional on-site potential terms.  Figure \ref{FigPotentialCorrection}(b) shows the energy spectrum of the $\mathcal{A}$ subspace for $N_c=15$ unit cells with $\rho_0=5\sigma$, distances $d'=3.6\sigma$, $d=4.5\sigma$, and  $U/t_s^\prime=20$ for (bi) an increasing potential correction $V$ in units of $V_\mathcal{A}$ and for (bii) the exactly compensated spectrum at $V_\mathcal{A}=(J_2'^2+J_3'^2)/U$. Figure \ref{FigPotentialCorrection}(bi) shows four bands with the outer ones presenting a larger dispersion than the inner ones, and also four Tamm-Shockley states for which the energy increases linearly with the potential correction $V$. As in the $\mathcal{B}$ subspace, the impurity states coincide in energy with the dispersive band due to the edge-bulk potential mismatch, but here two of them are located within the bulk. The fact that these two states, in the absence of any edge potential compensation, naturally appear within a bulk continuum while remaining localized at the edge, suggests that they may be regarded as bound states in the continuum \cite{Hsu2016}. Once the on-site potential mismatch is exactly compensated, introducing $V_\mathcal{A}$, the single-particle [Eq.~(\ref{EqKineticHamiltonianHubbard})] and two-particle [Eq.~(\ref{EqEffectiveA})] models become completely analogous. Then, one can apply the single-particle basis rotation (\ref{EqBasisChange}) for the $\mathcal{A}$ bound states that transforms the system into two decoupled SSH chains with the following renormalized couplings
\begin{equation}
	\begin{aligned}
		t_s^\mathcal{A}&=2\dfrac{J_2^2+J_3^2}{U} \qquad t_s^{\prime\mathcal{A}}&=2\dfrac{J_2'^2+J_3'^2}{U}\\
		t_a^\mathcal{A}&=2\dfrac{J_2^2-J_3^2}{U} \qquad t_a^{\prime\mathcal{A}}&=2\dfrac{J_2'^2-J_3'^2}{U}.
	\end{aligned}
\end{equation} 
The spectrum in Fig. \ref{FigPotentialCorrection}(bii) is equivalent to the single-particle one shown in Fig.~\ref{FigSpectrumSingleParticle} with renormalized and shifted energies. The outer bands, which show a greater dispersion, belong to the symmetric chain, while the inner ones belong to the antisymmetric chain. As both models are in the same topological phase, both lattices contribute with two topologically-protected edge states.

\section{Conclusion}\label{SecConclusion}
Here we have studied a system of one or two bosons loaded into states with OAM $l=1$ in a lattice of rings, with alternating distances $d$ and $d'$. By selecting the states with a given OAM $l$, each site of the lattice presents two internal states given by the two circulations $+$ and $-$. At the single-particle level, this system presents non-trivial topological characteristics, that can be properly analyzed by resolving the exchange symmetry between the circulations $+$ and $-$. This leads to two decoupled SSH chains whose associated Zak phases determine the topological phase of the system. We analyze the parameter space in terms of the distances $d$ and $d'$, finding that both chains are always in the same topological phase but show different dispersion in their bands. Thus, the system can present four topologically-protected edge states. 

Secondly, we study the case of two bosons with on-site interactions, which generate a rich landscape of doublon bands and edge bound states. In these bands, the two-particles occupy the same site, and we analyze them analytically in two limits. In the dimerized limit, one can reduce the system to a single strong-link. The eigenvalues of the associated Hamiltonian accurately predict the overall energy dependence of the doublon bands obtained using exact diagonalization in the nearly dimerized regime. The strong-link eigenvectors are analyzed in the strong interactions limit, where we find that they form three distinct groups. They either tend to states where the two  particles are in distinct sites, with energy zero, or to the same site in the same circulation (with energy $U$), or opposite circulations (energy $2U$). Additionally, we show that the avoided crossings between the doublon bands and the bands of extended states can only arise between eigenstates of the same exchange parity sector. In order to be able to capture the subspaces created by these doublon states away from the dimerized limit, we consider the strong interactions limit using second-order perturbation theory. The two doublon subspaces are well separated in energy and thus can be studied independently. We find effective models that map to an SSH model and a Creutz-like model with a bulk-edge on-site potential mismatch. We show how this mismatch can be corrected by introducing a potential at the edge sites, thus recovering the chiral symmetry that topologically protects the doublon edge states. The effective models are benchmarked using exact diagonalization.

\section{Acknowledgments}
E.N. is grateful to David Marin for helpful discussions. E.N., V.A., and J.M. acknowledge support through MCIN/AEI/ 10.13039/501100011033 Grant No. PID2020-118153GB-I00 and the Catalan Government (Contract No. SGR2021-00138). E.N. acknowledges financial support from MCIN/AEI/ 10.13039/501100011033 Contract No. PRE2018-085815 and from COST through Action CA16221. A.M.M. and R.G.D. acknowledge financial support from the Portuguese Institute for Nanostructures, Nanomodelling and Nanofabrication (i3N) through Projects No. UIDB/50025/2020, No. UIDP/50025/2020, and No. LA/P/0037/2020, and funding from FCT–Portuguese Foundation for Science and Technology through Project No. PTDC/FISMAC/29291/2017. A.M.M. acknowledges financial support from the FCT through the work Contract No. CDL-CTTRI147-ARH/2018 and from i3N through the work Contract No. CDL-CTTRI-46-SGRH/2022.

\onecolumngrid
\section*{Appendix}
The eigenvectors of the strong-link Hamiltonian discussed in Sec.~\ref{SecStrongLink} are given in Table \ref{TableEigenvectors} below. These correspond to the eigenvalues in Eq.~(\ref{EqEigenvaluesStrongLink}), where we consider the limit of large distances, $d\gg1$ and $J_3=J_2=J$. We have defined the factors 
\begin{equation}\label{EqFactorsEigenvectors}
	A^{+(-)}=\frac{U \varpm \sqrt{U^2+16 J^2}}{2\sqrt{2}J}, \qquad V=\dfrac{U}{2\sqrt{2}J},
\end{equation}
and the norms of the eigenvectors $\tilde{v}_{7,8,9,10}$ take the following simple forms
\begin{equation}
	\begin{array}{l}
		\|\tilde{v}_7\| = \|\tilde{v}_8\| =(8+4V^2 + 4V\sqrt{V^2 +2})^{1/2} \\[2mm]
		\|\tilde{v}_9\| = \|\tilde{v}_{10}\| = (8+4V^2 - 4V\sqrt{V^2 +2})^{1/2}.
	\end{array}
\end{equation}

\begin{table*}[h]
	\setlength\tabcolsep{1.1mm}
	\begin{center}
		\begin{tabular}{c c c c c c c c c c c}
			\hline\hline\rule{0pt}{12pt}
			$v_n$ &
			$\left|A_{m}^+ A_{m}^+\right\rangle$&
			$\left|A_{m}^+ A_{m}^-\right\rangle$&
			$\left|A_{m}^+ B_{m}^+\right\rangle$&
			$\left|A_{m}^+ B_{m}^-\right\rangle$&
			$\left|A_{m}^- A_{m}^-\right\rangle$&
			$\left|A_{m}^- B_{m}^+\right\rangle$&
			$\left|A_{m}^- B_{m}^-\right\rangle$&
			$\left|B_{m}^+ B_{m}^+\right\rangle$&
			$\left|B_{m}^+ B_{m}^-\right\rangle$&
			$\left|B_{m}^- B_{m}^-\right\rangle$\\[1mm]\hline
			$v_1$&$0$&$0$&$\frac{1}{2}$&$-\frac{1}{2}$&$0$&$-\frac{1}{2}$&$\frac{1}{2}$&$0$&$0$&$0$\\
			$v_2$&$-\frac{1}{2}$&$0$&$0$&$0$&$-\frac{1}{2}$&$0$&$0$&$\frac{1}{2}$&$0$&$\frac{1}{2}$\\
			$v_3$&$0$&$-\frac{1}{\sqrt{2}}$&$0$&$0$&$0$&$0$&$0$&$0$&$\frac{1}{\sqrt{2}}$&$0$\\
			$\tilde{v}_4$&$1$&$\sqrt{2}(E_4^2-E_4V-1)$&$E_4-V$&$E_4-V$&$1$&$E_4-V$&$E_4-V$&$1$&$\sqrt{2}(E_4^2-E_4V-1)$&$1$\\
			$\tilde{v}_5$&$1$&$\sqrt{2}(E_5^2-E_5V-1)$&$E_5-V$&$E_5-V$&$1$&$E_5-V$&$E_5-V$&$1$&$\sqrt{2}(E_5^2-E_5V-1)$&$1$\\
			$\tilde{v}_6$&$1$&$\sqrt{2}(E_6^2-E_6V-1)$&$E_6-V$&$E_6-V$&$1$&$E_6-V$&$E_6-V$&$1$&$\sqrt{2}(E_6^2-E_6V-1)$&$1$\\
			$\tilde{v}_7$&$1$&$0$&$0$&$-A^+$&$-1$&$A^+$&$0$&$-1$&$0$&$1$\\
			$\tilde{v}_8$&$1$&$0$&$-A^+$&$0$&$-1$&$0$&$A^+$&$1$&$0$&$-1$\\
			$\tilde{v}_9$&$1$&$0$&$0$&$-A^-$&$-1$&$A^-$&$0$&$-1$&$0$&$1$\\
			$\tilde{v}_{10}$&$1$&$0$&$-A^-$&$0$&$-1$&$0$&$A^-$&$1$&$0$&$-1$
		\end{tabular}
	\end{center}
	\caption{Eigenvectors $v_n$ of the strong-link Hamiltonian $\hat{\mathcal{H}}_{SL}$ in the limit of large distances $d \gg 1$, for which $J_2=J_3=J$. The normalized eigenvectors are $v_n=\tilde{v}_n/\|\tilde{v}_n\|$ and we use the factors $A^{+(-)}$ and $V$ defined in Eq.~(\ref{EqFactorsEigenvectors}). }
	\label{TableEigenvectors}
\end{table*}

\twocolumngrid

\end{document}